\documentclass{article}

\usepackage{mathtools}
\usepackage{amssymb}
\usepackage{dsfont}
\usepackage{slashed}
\usepackage{fullpage}
\usepackage{xcolor}

\usepackage[colorlinks=true,linkcolor=blue,citecolor=magenta,linktocpage=true]{hyperref}
\usepackage{titlesec}
\titleformat*{\section}{\normalsize\bfseries}
\titleformat*{\subsection}{\normalsize\bfseries}
\titleformat*{\subsubsection}{\normalsize\bfseries}
\usepackage{cite}
\usepackage{cancel}

\DeclareMathAlphabet{\bbvar}{U}{BOONDOX-ds}{m}{n}

\makeatletter
\renewcommand{\@dotsep}{10000}
\makeatother

\newcommand{\cB}{{\mathcal B}}

\newcommand{\cL}{{\mathcal L}}
\newcommand{\cH}{{\mathcal H}}
\newcommand{\cM}{{\mathcal M}}

\newcommand{\cR}{{\mathcal R}}

\newcommand{\cV}{{\mathcal V}}

\newcommand{\cS}{{\mathcal S}}

\newcommand{\SL}{\mathrm{SL}}

\newcommand{\be}{\begin{equation}}
\newcommand{\ee}{\end{equation}}
\newcommand{\beq}{\begin{eqnarray}}
\newcommand{\eeq}{\end{eqnarray}}
\newcommand{\bes}{\begin{eqnarray}}
\newcommand{\ees}{\end{eqnarray}}

\def\rd{\textrm{d}}

\def\beq{\begin{eqnarray}}
\def\eeq{\end{eqnarray}}
\def\be{\begin{equation}}
\def\ee{\end{equation}}

\def\r2A{z}

\usepackage[toc,page]{appendix}
\numberwithin{equation}{section}

\usepackage{url}

 \usepackage{amsmath}
\usepackage{graphicx}
\usepackage{epstopdf}
\usepackage{float}
\usepackage{hyperref}
\usepackage{color}
\usepackage[T1]{fontenc}
\usepackage[utf8]{inputenc}
\usepackage[toc,page]{appendix}
\usepackage[normalem]{ulem}
\usepackage{caption}
\usepackage{subcaption}

\usepackage{tikz}
\usetikzlibrary{shapes,positioning,shadows,graphs.standard,automata,arrows}
\usepackage{pgfplots}
\pgfplotsset{compat=newest, width=2.669cm, height=2.669cm, scale only axis=true,enlargelimits=false}
\pgfplotsset{tick label style={font=\tiny}}
\pgfplotsset{every major tick/.append style={major tick length=3pt}}
\pgfplotsset{every minor tick/.append style={minor tick length=1.5pt}}
\usepgfplotslibrary{groupplots}
\usepackage{caption}
\usepackage[font=small,labelfont=bf,justification=justified,format=plain]{caption}

\providecommand{\renewoperator}[3]{%
\renewcommand*{#1}{\mathop{#2}#3}}

\makeatletter
\providecommand*{\diff}%
{\@ifnextchar^{\DIfF}{\DIfF^{}}}
\def\DIfF^#1{%
\mathop{\mathrm{\mathstrut d}}%
\nolimits^{#1}\gobblespace}
\def\gobblespace{%
\futurelet\diffarg\opspace}
\def\opspace{%
\let\DiffSpace\!%
\ifx\diffarg(%
\let\DiffSpace\relax
\else
\ifx\diffarg[%
\let\DiffSpace\relax
\else
\ifx\diffarg\{%
\let\DiffSpace\relax
\fi\fi\fi\DiffSpace}

\renewoperator{\Re}{\mathrm{Re}}{\nolimits}
\renewoperator{\Im}{\mathrm{Im}}{\nolimits}

\usepackage{bm}

\newcommand{\ba}{\begin{eqnarray}}
\newcommand{\ea}{\end{eqnarray}}

\newcommand{\beqa}{\begin{eqnarray}}
\newcommand{\eeqa}{\end{eqnarray}}

\newtheorem{theorem}{Theorem}

\begin{document}

\title{\Large{\sffamily Dynamical axisymmetric compact objects \\ in General Relativity}}

\author{\sffamily Jibril Ben Achour\;$^{1,2}$, Adolfo Cisterna\;$^{3}$ and Mokhtar Hassaine\;$^{4}$} 
\date{\small{\textit{$^{1}$ Arnold Sommerfeld Center for Theoretical Physics, Munich, Germany \\
$^{2}$ Univ Lyon, CNRS, ENS de Lyon, Laboratoire de Physique (LPENSL), Lyon, France\\
$^{3}$ Sede Esmeralda, Universidad de Tarapac{\'a}, Avenida Luis Emilio Recabarren 2477, Iquique, Chile\\
${4}$ Instituto de Matem\'{a}ticas,
Universidad de Talca, Casilla 747, Talca, Chile\\
}}}

\maketitle

\begin{abstract} 
The search for exact solutions describing asymptotically FLRW compact objects in General Relativity remains a challenging problem. Progress has largely been limited to the spherically symmetric case, with notable exceptions such as the Kerr–de Sitter and Thakurta solutions. In this work, we present two new results that advance the description of axisymmetric compact objects embedded in a cosmological background.
First, we introduce a new solution-generating technique that allows for the construction of nonstationary, axisymmetric solutions of the self-interacting Einstein-scalar system. Using this method, we obtain the first exact solution that can describe a dynamical axisymmetric compact object in a FLRW cosmology.
We then outline how a detailed analysis of its properties, particularly dynamical trapping (or anti-trapping) horizons, can be carried out. For this purpose, we employ the mean curvature vector (MCV), which provides a natural extension of the Kodama vector beyond spherical symmetry. The norm of the MCV defines a foliation-independent, though embedding-dependent, quantity that can be used to identify trapped, anti-trapped, and untrapped regions, and to characterise the causal structure of the geometry without relying on specific symmetry assumptions.
The embedding dependence must be treated carefully, as it determines the extent to which the analysis can be performed analytically while minimising the use of numerical methods.
Overall, the solution-generating approach and the associated analysis tools offer a framework to further investigate dynamical axisymmetric compact objects, including black holes in cosmological settings and scenarios involving dynamical scalar accretion.

\end{abstract}

\thispagestyle{empty}
\newpage
\setcounter{page}{1}

\hrule
\tableofcontents
\vspace{0.7cm}
\hrule

\newpage

\section{Introduction}

Primordial black holes (PBHs) may have formed in the early Universe through a variety of mechanisms. Although studies of these compact objects without stellar origin began several decades ago \cite{Carr:2024nlv}, PBHs have triggered a renewed interest as potential candidates for the still elusive dark matter \cite{Villanueva-Domingo:2021spv}. Their possible cosmological role depends on several factors, including their abundance, formation mechanisms, evaporation rate, and their capacity to accrete matter, merge, or cluster. These properties can influence cosmological observables in multiple ways. In addition, PBHs could contribute to the early formation of structures at high redshift, as suggested by recent JWST observations \cite{Castellano:2022ikm}.
Constraints on PBHs span a wide mass range and often rely on assumptions about their evaporation\footnote{In particular, most analyses implicitly assume that dynamical PBHs obey the same qualitative mass-loss relation as asymptotically flat Schwarzschild black holes, as predicted by semiclassical Hawking radiation, where the black hole temperature scales inversely with the mass, $T_{\text{BH}} \propto M^{-1}$ \cite{Mallett:1986hr, Koberlein:1995up, Nielsen:2005af}. However, it is well known that dynamical horizons are characterized by a different notion of temperature, which in turn implies a more intricate evaporation process \cite{Senovilla:2014ika}.}. Based on this, limits have been derived from the potential emission of high-energy particles near the end of their lifetime \cite{Cheek:2021odj, Cheek:2022mmy, Korwar:2023kpy, Perez-Gonzalez:2025try, DelaTorreLuque:2024qms, Klipfel:2025bvh}.
Future gravitational-wave observations are also expected to probe PBHs, either through their contribution to the stochastic background, the production of induced second-order gravitational waves, or the detection of extreme mass-ratio inspirals with LISA \cite{LISACosmologyWorkingGroup:2023njw, Domenech:2024cjn, Papanikolaou:2020qtd, DeLuca:2025uov, Crescimbeni:2024cwh, Gross:2025hia, Begnoni:2025aqc}. For an overview of the different observational prospects, see \cite{Riotto:2024ayo}.

On the theoretical side, research has mainly concentrated on two directions: (i) developing and testing criteria for PBH formation \cite{Shibata:1999zs, Harada:2015yda, Escriva:2019phb, Kalaja:2019uju, Musco:2020jjb, Escriva:2020tak, Harada:2023ffo, Harada:2024trx, Kehagias:2024kgk}, and (ii) estimating PBH abundances from the statistical properties of large cosmological perturbations in different scenarios \cite{Ferrante:2022mui, Animali:2022otk, Gow:2022jfb}. Reviews can be found in \cite{Escriva:2021aeh, Escriva:2022duf, Yoo:2022mzl}.
In contrast, the search for exact solutions of the Einstein field equations representing asymptotically FLRW black holes, initiated decades ago \cite{McVittie, Straus, Tolman}, has remained relatively limited until more recent work \cite{Lasota, Thakurta, Husain:1994uj, Fonarev:1994xq, Kothawala:2004fy, Gibbons:2009dr, Akcay:2010vt, Mello:2016irl, Babichev:2018ubo, Xavier:2021chn, Croker:2021duf, Heydari:2021gea, Sato:2022yto, Babichev:2023mgk, Tang:2024cfy, Rasulian:2025jpp}, and has been the subject of renewed debate \cite{Kobakhidze:2021rsh, Hutsi:2021nvs, Boehm:2021kzq, Harada:2021xze, Maciel:2024eys} (see \cite{Faraoni:2013aba, Faraoni:2018xwo} for reviews).
Exact solutions describing compact objects embedded in cosmology are valuable in several respects. They provide analytic benchmarks for testing numerical methods, and they can uncover subtle non-linear effects, particularly regarding horizon formation and dynamics, that may be difficult to identify numerically. Moreover, they are essential for developing perturbative frameworks to study gravitational radiation and the evaporation of these dynamical compact objects.

A main challenge in deriving exact solutions for PBHs is that, by definition, they correspond to asymptotically FLRW and non-vacuum spacetimes, making them inherently dynamical. This represents a departure from the more familiar setting of vacuum, asymptotically flat geometries, with important consequences such as the loss of a timelike Killing vector.

Early attempts to construct exact inhomogeneous solutions embedded in cosmology go back to McVittie and Tolman \cite{McVittie, Tolman}, and were later extended by models such as the Einstein–Straus vacuole (the “Swiss-cheese” solution) and the LTB family \cite{Straus, Bondi:1947fta}. These works were originally motivated by the question of how cosmic expansion affects quasi-local objects like stars and black holes. Addressing this issue is nontrivial because, in gravity, the separation between long and short scales is far from obvious. While a perturbative scheme with a fixed background can impose some control, fully nonlinear exact solutions inherently involve scale mixing.
This difficulty is well illustrated by the nontrivial local effects that arise when introducing a cosmological constant in different spacetimes \cite{Adams:1999pr, Faraoni:2007es, Carrera:2008pi, Nandra:2011ug}, as well as by the subtleties it introduces in defining and characterizing gravitational radiation in asymptotically de Sitter geometries, a topic currently under investigation \cite{Ashtekar:2015ooa, Ashtekar:2015lxa, Ashtekar:2017dlf, Compere:2023ktn, Fernandez-Alvarez:2025qqx, Fernandez-Alvarez:2024bkf, Fernandez-Alvarez:2021uvz}.

The study of asymptotically FLRW and fully dynamical black holes faces two main challenges.
First, explicitly solving the nonlinear Einstein equations in a non-stationary setting is notoriously difficult, as reflected by the limited number of known exact solutions (see \cite{Faraoni:2013aba, Faraoni:2018xwo} for reviews). A common strategy is to conformally rescale a stationary solution, such as Schwarzschild, with a time-dependent scale factor and then determine the energy–momentum tensor implied by the field equations. However, this does not truly solve the dynamics. Such constructions often introduce pathologies in the matter sector, particularly violations of energy conditions, casting doubt on their physical relevance. The Thakurta solution \cite{Thakurta}, frequently used in recent PBH studies, provides a typical example. Similarly, Swiss-cheese–type models, while mathematically consistent at the background level, are unstable once perturbations are included.
In contrast, approaches where the metric and matter content are solved together have advanced more slowly; however, some solution-generating techniques exist, especially in the self-interacting Einstein–Scalar system \cite{Husain:1994uj, Fonarev:1994xq}. These yield genuine dynamical black (or white) holes embedded in an inflationary universe.  
Such solutions have so far remained largely confined to the exact solutions community. A central aim of our work is to extend their construction to broader geometrical settings. 

The second difficulty arises even when an exact solution is available: analysing its causal structure and physical interpretation is highly nontrivial. This is exemplified by the McVittie solution, whose black hole status was debated for decades and only clarified relatively recently \cite{Nolan:1998xs, Nolan:1999wf, Nolan:1999kk, Kaloper:2010ec}. A similar discussion surrounds the Thakurta solution \cite{Kobakhidze:2021rsh, Hutsi:2021nvs, Boehm:2021kzq, Harada:2021xze, Maciel:2024eys}. The difficulty stems from the absence of a timelike Killing vector in dynamical settings, meaning that the standard notion of a Killing horizon no longer applies. While well-defined frameworks for dynamical horizons out of equilibrium have been developed \cite{Ashtekar:2025wnu, Ashtekar:2002ag, Ashtekar:2003hk, Ashtekar:2004cn, Ashtekar:2013qta}, concretely identifying horizons in a given solution remains challenging. In particular, quasi-local (anti-)trapping horizons are defined through the expansions of null rays, expansions that are foliation dependent \cite{Wald:1991zz, Schnetter:2005ea, Faraoni:2016xgy, Dotti:2023elh, Dotti:2025npw}.

In this context, the spherically symmetric case is special. As first shown by Kodama \cite{Kodama:1979vn}, any dynamical spherically symmetric spacetime admits a divergence-free vector field, now known as the Kodama vector\footnote{This vector arises from the existence of a rank-2 Killing–Yano tensor present in any dynamical spherically symmetric geometry \cite{Kinoshita:2024wyr}, and can thus be viewed as a manifestation of a hidden symmetry in this class of spacetimes.}. The Kodama vector has three main properties. First, although it does not satisfy the Killing equation, it defines a preferred notion of time (i.e. a preferred slicing), which greatly simplifies the Einstein field equations \cite{Racz:2005pm, Csizmadia:2009dm}. Second, it leads to conserved currents and associated charges that play an important role in the thermodynamics of these geometries \cite{Abreu:2010ru, Achour:2021lqq}. Third, and most relevant for our purposes, the Kodama vector becomes null on a trapping (or anti-trapping) horizon, in analogy with the Killing vector becoming null on the Killing horizon of a stationary black hole. More generally, the Kodama vector is timelike in untrapped regions and spacelike in trapped (or anti-trapped) regions. Since its norm is foliation independent, it provides an invariant way to characterise the causal structure and, in particular, the dynamical horizons of spherically symmetric solutions.
The extension of the Kodama vector beyond spherical symmetry is known as the mean curvature vector (MCV)\footnote{Interestingly, the search for a Kodama-like vector in axisymmetry has been pursued despite the existence of the MCV. See for instance \cite{Kinoshita:2021qsv, Dorau:2024zyi}.}. Despite a few applications to track the horizons of evaporating black holes \cite{Senovilla:2014ika, Dotti:2023elh, Dotti:2025npw} and, in particular, those of the Kerr–Vaidya geometry \cite{Dahal:2021vjr, Dahal:2023zks}, the use of the MCV has remained limited to the community working in mathematical general relativity \cite{Anco:2004bb}\footnote{An early work in which the MCV was used to describe trapped surfaces can be found in \cite{Mars:2003ud}.} The MCV is a key object to characterise the structure of (dynamical) horizons of axisymmetric and dynamical compact objects.

Thus, a second goal of this work is to review and advertise this use of the mean-curvature vector for an arbitrary spacetime and to show its relevance (and limitations) when exploring the structure of axisymmetric compact objects embedded in cosmology.
Concretely, we develop a new solution-generating method for constructing asymptotically FLRW, axisymmetric solutions of the self-interacting Einstein–Scalar system. Using this method, we construct and focus on one explicit example and assess up to which extent the MCV can be used to analyse the dynamical horizons of this non-stationary, axisymmetric geometry. 
To our knowledge, aside from the Kerr–de Sitter solution and the rotating Thakurta model (the latter based on a perfect fluid but exhibiting pathologies), this represents the first exact, non-vacuum, asymptotically FLRW, axisymmetric solution of the Einstein–Scalar system with a compact source.
Interestingly, the search for such axisymmetric solutions motivated our investigation into a generalisation of the Kodama vector, an object that, as it turned out, was already provided by Anco’s construction \cite{Anco:2004bb}.   

This work is organised as follows. In Section~\ref{SEC1}, we review the properties of the Kodama vector in spherical symmetry. Building on Anco’s work \cite{Anco:2004bb}, we then introduce the MCV, defined in (\ref{MCV}–\ref{MCVD}), and show how it extends the key properties of the Kodama construction.
Section~\ref{SEC2} presents the new solution-generating method. We begin by reviewing the Buchdahl and Fonarev techniques, which respectively allow the construction of static axisymmetric and dynamical spherically symmetric solutions from a static spherical seed. We then show how these approaches can be combined to obtain dynamical, axisymmetric solutions of the self-interacting Einstein–Scalar system, and we provide a complete proof of the method.
In Section~\ref{SEC3}, we apply this framework to construct the first dynamical axisymmetric solution of the system, given in (\ref{METSOL}–\ref{SCSOL}). We first analyse the seed, the Zipoy–Voorhees solution, and its properties; then we present its dynamical extension. 
On this dynamical geometry, we first identify its asymptotic FLRW behaviour. We then describe how the analysis of its dynamical horizons can be carried out, focusing on the identification of trapped and anti-trapped regions, the types of horizons that may form, and the extent to which their properties can be determined analytically using the MCV. We also discuss a main obstruction that necessitates the use of numerical methods when defining the embedding on which the MCV is constructed. Finally, we indicate which classes of axisymmetric geometries can avoid this issue, allowing for a more fully analytic study of the dynamical horizons of axisymmetric dynamical spacetimes.
We conclude with a discussion of the perspectives opened by these results, for the broader search for exact, physically relevant dynamical solutions.

\section{Generalized Kodama vector beyond spherical symmetry}

\label{SEC1}

In this section, we review the definition of the mean curvature vector (and its dual), as introduced in \cite{Anco:2004bb}, and discuss how it naturally generalises the Kodama vector for spherically symmetric spacetimes, a fact that was already noticed in \cite{Senovilla:2014ika}. In particular, we review how it provides a foliation-independent method to locate trapping and anti-trapping dynamical horizons in any spacetime, without assuming specific symmetries. The main goal of this section is to highlight its key role in studying dynamical compact objects beyond spherical symmetry. As a starting point, we first recall the definition and key properties of the Kodama vector.

\subsection{Review of the Kodama vector and its properties}

Consider a time-dependent, spherically symmetric spacetime. Without loss of generality, its metric can be expressed as
\be
\label{metSS}
\rd s^2 = - e^{-2\Phi} f \rd t^2 + \frac{\rd r^2}{f} + R^2 \rd \Omega^2,
\ee
with the corresponding fields $\Phi:= \Phi(t,r)$, $f:= f(t,r)$ and $R:= R(t,r)$.
We parametrize the function $f(t,r)$ in the standard form 
\be
f(t,r) = 1 - \frac{2m(t,r)}{r},
\ee
where $m(t,r)$ is the Misner–Sharp mass \cite{Misner:1964je}. In the gauge $R(t,r) = r$, the horizon can be located by setting $g^{rr} = 0$, which corresponds to the hypersurface equation $r_H = 2m(t_H, r_H)$.

Any such spacetime possesses a hidden symmetry generated by a Killing–Yano tensor, i.e., an antisymmetric rank-2 tensor satisfying $\nabla_{(\mu} Y_{\nu)\alpha} = 0$, which can be expressed in the form
\be
Y_{\mu\nu} \rd x^{\mu} \rd x^{\nu} = R^3(t,r) \sin{\theta} \rd \theta \wedge \rd \varphi .
\ee
Now, any Killing–Yano 2-form naturally defines a dual vector field, given by
\begin{align}
K^{\mu} \partial_{\mu}& = \frac{1}{2} \varepsilon^{\mu\alpha\beta\gamma} \nabla_{\alpha} Y_{\beta\gamma}  \partial_{\mu} ,
\end{align}
which by construction, satisfies $\nabla_{\mu} K^{\mu} = 0$, thereby providing a locally conserved current even in fully inhomogeneous and dynamical spacetimes.
A straightforward computation shows that the only nonvanishing components of this vector are
\begin{align}
K^t & =  \frac{1}{2} \varepsilon^{tr\theta\varphi} \left( Y'_{\theta\varphi} - \Gamma^{\sigma}{}_{r\theta} Y_{\sigma\varphi} - \Gamma^{\sigma}{}_{r\varphi} Y_{\sigma\theta}  \right) =  { -\frac{3}{2}e^{\Phi}}  R' , \\ 
K^r & =  \frac{1}{2} \varepsilon^{rt \theta\varphi} \left( \dot{Y}_{\theta\varphi} - \Gamma^{\sigma}{}_{t\theta} Y_{\sigma\varphi} - \Gamma^{\sigma}{}_{t\varphi} Y_{\sigma\theta}  \right) =  { \frac{3}{2}e^{\Phi}}  \dot{R} ,
\end{align}
where, we recall that $\varepsilon^{\mu\alpha\beta\gamma} = \epsilon^{\mu\alpha\beta\gamma}/\sqrt{|g|}$, with $\sqrt{|g|} = e^{-\Phi} R^2 \sin\theta$.
Finally, the dual vector and its norm read
\be
K^{\mu} \partial_{\mu} = { \frac{3}{2}} e^{\Phi} \left( R' \partial_t + \dot{R} \partial_r \right) ,\qquad{ K^{\mu}K_{\mu} =\frac{9}{4}\left( e^{2\Phi} \frac{\dot{R}^2}{f} - f R'^2\right)}.
\ee
This precisely reproduces the form of the Kodama vector for a general spherically symmetric spacetime \cite{Kodama:1979vn, Abreu:2010ru}. Consequently, the existence of the Kodama current follows directly from the Killing–Yano symmetry of the underlying spherically symmetric geometry. See \cite{Kinoshita:2024wyr} for a recent discussion of this point.

For dynamical spherically symmetric spacetimes, the importance of the Kodama vector cannot be overemphasised. It possesses four crucial properties. 
\begin{itemize}
\item First, and most relevant for our purposes, the Kodama vector becomes null on the apparent horizons of any dynamical spherically symmetric spacetime. To see this, consider the gauge choice $R(t,r) = r$ in (\ref{metSS}). Then, one has
\begin{equation}
{ K^{\mu}K_{\mu}=\frac{9}{4}\left( \frac{2 m(t,r)}{r} - 1 \right)},
\end{equation}
which vanishes at the dynamical horizon $r_H = 2 m(t, r_H)$. Moreover, one sees that it is timelike for $r > r_H$ and spacelike for $r < r_H$. The norm of the Kodama vector thus provides an efficient, coordinate-independent tool to locate apparent horizons and to study the causal nature of different regions.
However, while the sign of the norm allows us to distinguish untrapped from (anti-)trapped regions, it does not distinguish between trapped and anti-trapped regions. As we shall see in the next section, the MCV provides a generalisation of this key object to an arbitrary spacetime without imposing any symmetry restrictions.
\item Second, it singles out a preferred time direction in any dynamical spherically symmetric spacetime, which can then be used to significantly simplify the evolution and constraint equations for this class of geometries \cite{Abreu:2010ru, Racz:2005pm, Csizmadia:2009dm}.
\item Third, the Kodama vector is divergence-free, i.e., $\nabla_{\mu} K^{\mu} = 0$, and therefore defines a conserved current valid for any dynamical spherically symmetric geometry. This property, in turns, allows the construction of additional conserved currents, as studied in \cite{Abreu:2010ru}. The simplest and most direct example is
\begin{equation}
J^{\mu} = G^{\mu\nu} K_{\nu},
\end{equation}
being $G^{\mu\nu}$ the Einstein tensor, whose conservation follows from the contracted Bianchi identity. Interestingly, it was shown in \cite{Achour:2021lqq} that, at least in the context of homogeneous cosmological geometries, the charges associated with these currents form an $\SL(2,R)$ algebra, known as the complexifier-volume-Hamiltonian (CVH) algebra, providing a useful algebraic structure for this class of geometries. See \cite{BenAchour:2022fif, BenAchour:2023dgj, Achour:2021dtj, Geiller:2021jmg, BenAchour:2019ufa} for detailed investigations on this inbuilt SL(2,R) structure and its extensions in black hole and cosmology.
\item Finally, the Kodama vector was used by Hayward in \cite{Hayward:1997jp} to introduce a notion of dynamical surface gravity for spherically symmetric dynamical horizons. This generalization, known as the Hayward–Kodama surface gravity, has been studied extensively, both in the context of formulating the thermodynamical laws of dynamical black holes and in relation to the evaporation of non-stationary horizons \cite{Nielsen:2007ac, Pielahn:2011ra, Cai:2006rs, Cai:2008gw, Helou:2015yqa, Helou:2015zma, Helou:2016xyu, Faraoni:2016xgy, Pathak:2023nip}, for both black holes and cosmological horizons.
\end{itemize}
Having reviewed the key properties and advantages of the Kodama vector for (dynamical) spherically symmetric geometries, let us now present its generalization following Anco's work \cite{Anco:2004bb}.

\subsection{From the Kodama to the mean curvature vector}

Consider a closed region $\cV$ in a four-dimensional spacetime manifold $(\cM, g)$ for which the boundary $\partial \cV$ decomposes as $\partial \cV = \Sigma_i \cup \cB \cup \Sigma_f$, where $(\Sigma_{i,f}, h)$ are the spacelike hypersurfaces at initial and final times with induced metric $h$, and $(\cB, \gamma)$ is a timelike hypersurface with induced metric $\gamma$.

\subsubsection{Definitions}

Now, consider a constant-time spacelike hypersurface $\Sigma_t$. Its intersection with the timelike boundary $\cB$ defines a closed 2-surface $\cS = \Sigma_t \cap \cB$. In a cosmological context, this surface represents the celestial sphere. Let us introduce the unit normal vector $n_{\mu} \rd x^{\mu}$ to $\Sigma$, with $n^{\mu} n_{\mu} = -1$, which is future-pointing, and the unit normal vector $s_{\mu} \rd x^{\mu}$ to $\cB$, with $s^{\mu} s_{\mu} = +1$, which is inward-pointing and satisfies $g_{\mu\nu} n^{\mu} s^{\nu} = 0$. The metric on $\cS$ can be expressed as
\be
q_{\mu\nu} = g_{\mu\nu} + n_{\mu} n_{\nu} - s_{\mu} s_{\nu} ,
\ee
such that $q_{\mu\nu} n^{\mu} = q_{\mu\nu} s^{\mu} = 0$. Therefore, $q_{\mu}{}^{\nu}$ acts as a projector onto $\cS$ and one can define the covariant derivative on $\cS$ by $D_{\mu} = q_{\mu}{}^{\nu} \nabla_{\nu}$.

Now, since $\cS$ is a 2-surface in the four-dimensional spacetime $\cM$, its extrinsic curvature can be decomposed into contributions within the hypersurface $\Sigma$ and within the hypersurface $\cB$, which are respectively defined by
\begin{align}
K_{\mu\nu} (n)=  D_{\mu} n_{\nu}, \qquad K_{\mu\nu} (s)=  D_{\mu} s_{\nu} .
\end{align}
At this stage, to encode the bending of $\cS$ within $\cM$, one can define the mean curvature vector $H$ and its dual $H_{\perp}$ (following the notation introduced in \cite{Anco:2004bb}) as
\begin{align}
\label{MCV}
H^{\mu} \partial_{\mu} & = \frac{1}{2} \left( K(s) s^{\mu} \partial_{\mu} - K(n) n^{\mu}\partial_{\mu}  \right),  \\
\label{MCVD}
H_{\perp}^{\mu} \partial_{\mu} & = \frac{1}{2} \left( K(s) n^{\mu} \partial_{\mu} - K(n) s^{\mu}\partial_{\mu} \right) ,
\end{align}
where $K(n)$ and $K(s)$ are the traces of the extrinsic curvatures, i.e. $K(n) = D_{\mu} n^{\mu}$ and $K(s) = D_{\mu} s^{\mu}$.
Notice that, by construction, one automatically has $g_{\mu\nu} H^{\mu} H^{\nu}_{\perp} = 0$. The pair $(H, H_{\perp})$ is referred to as the mean curvature (orthogonal) frame of $S^{\perp}$.
Finally, the norm of both vectors is  given by
\begin{equation}
g_{\mu\nu}H^{\mu} H^{\nu} = - g_{\mu\nu} H^{\mu}_{\perp} H^{\nu}_{\perp} = \frac{K^2(s) - K^2(n)}{4},
\end{equation}
such that
\begin{align}
\label{norm}
 |H| = |H_{\perp}| = \frac{1}{2} \sqrt{K^2(s) - K^2(n)}.
\end{align}
An interesting feature of the pair $(H, H_{\perp})$ is that
\be
K(H_{\perp}) = D_{\mu} H^{\mu}_{\perp} =0, \qquad K(H) = D_{\mu} H^{\mu} = H^2 = H^2_{\perp}.
\ee 
Therefore, the dual mean curvature vector $H_{\perp}$ singles out the normal direction with respect to $\cS$ in which the surface has vanishing extrinsic curvature. It turns our that $H_{\perp}$ is a key object for constructing an invariant notion of quasi-local energy for the closed 2-surface $\cS$. Moreover, $H_{\perp}$ provides a natural generalization of the Kodama vector beyond spherical symmetry. As we will show, it shares similar properties and can be used to track the presence of horizons within the geometry.

\subsubsection{Localizing the horizons}

Instead of the basis $(n,s)$ of $T_p(\cS^{\perp})$, one can introduce a pair of outward- and inward-pointing null vectors $(\ell_+, \ell_-)$ such that $g_{\mu\nu} \ell_{+}^{\mu} \ell_{-}^{\nu} = -1$, which are related to the pair $(n,s)$ as
\be
\ell_{+}^{\mu} \partial_{\mu} =  \frac{1}{\sqrt{2}}(n^{\mu}+s^{\mu}) \partial_{\mu}, \qquad \ell_{-}^{\mu} \partial_{\mu} = \frac{1}{\sqrt{2}} (n^{\mu} -s^{\mu}) \partial_{\mu}, 
\ee
 where $\ell_{+}$ denotes the outward-pointing null vector, and $\ell_{-}$ denotes the inward-pointing null vector.
The metric on $\cS$ is now written as
\be
q_{\mu\nu} = g_{\mu\nu} + \ell^{+}_{\mu} \ell^{-}_{\nu} + \ell^{+}_{\nu} \ell^{-}_{\mu}.
\ee
Once again, defining the projected derivative on $\cS$ as $D_{\mu} = q_{\mu}{}^{\nu} \nabla_{\nu}$, one can construct the extrinsic curvature of $\cS$ along each null direction
\begin{align}
K_{\mu\nu} (\ell^{+}) &= D_{\mu} \ell^{+}_{\nu}, \qquad
K_{\mu\nu} (\ell^{-}) = D_{\mu} \ell^{-}_{\nu}.
\end{align}
It can be noted that their traces correspond precisely to the expansions of the respective null vectors, namely
\be
K(\ell_{+}) = D_{\mu} \ell_{+}^{\mu} = \theta_{+}, \qquad K(\ell_{-}) = D_{\mu} \ell_{-}^{\mu} = \theta_{-}.
\ee
Following the same construction as above, the mean curvature vector and its dual can be expressed as
\begin{align}
H^{\mu} \partial_{\mu} & = -\frac{1}{2} \left(  \theta_{+} \ell_{-}^{\mu} \partial_{\mu} + \theta_{-} \ell_{+}^{\mu} \partial_{\mu} \right), \\
H_{\perp}^{\mu} \partial_{\mu} & =\;\; \; \frac{1}{2} \left(  \theta_{+} \ell_{-}^{\mu} \partial_{\mu} - \theta_{-} \ell^{\mu}_{+} \partial_{\mu} \right) ,
\end{align}
such that the corresponding norms yield
\begin{align}
\label{exp}
{H^2 = - H^2_{\perp} =- \frac{\theta_{+} \theta_{-}}{2}}
\end{align}
Being directly related to the expansions of the outward and inward null vectors, they can be used to track the presence of (anti-)trapping horizons and, consequently, (anti-)trapped regions, just as the Kodama vector does in spherical symmetry. It is important to emphasize that, in general, different choices of tetrad correspond to different null vectors and therefore yield different expansions with distinct “zero-loci.” Only in very special circumstances does the vanishing of an expansion correspond to a genuine geometric surface, such as a horizon. Since each null congruence intersects a horizon differently—or may fail to intersect it altogether—its transverse deformations and associated expansions generally exhibit distinct behaviors.
The expansion of a null congruence acquires a geometrically meaningful and invariant interpretation only under a specific condition: both null vectors must be orthogonal to a fixed spacelike two-surface. This is precisely the framework underlying trapped surfaces, marginally trapped surfaces, marginally outer trapped surfaces, and quasi-local horizon theory.
This is exactly the situation considered here. The null tetrad used to construct the MCV is defined by null vectors that are orthogonal to a physically meaningful two-dimensional surface. As a result, the corresponding zero-loci faithfully reproduce the location of trapped and anti-trapped surfaces in a fully coordinate-independent manner. 

Now, for clarity, let us briefly recall the different definitions of horizons and examine how the norm of the vectors $(H, H_{\perp})$ behaves in each case.
\begin{itemize}
\item First, a closed $2$-surface is said to be untrapped if the expansions of ingoing and outgoing light rays evaluated on that surface satisfy $\theta_{+}>0$ and $\theta_{-}<0$. Then, the closed $2$-surface where $\theta_{+} = 0$ and $\theta_{-} < 0$ defines future trapping horizons (corresponding either to black holes or contracting cosmological horizons), whereas the closed $2$-surface where $\theta_{-} = 0$ and $\theta_{+} > 0$ defines past anti-trapping horizons (corresponding to expanding cosmological horizons or white holes). They correspond to the trapped or anti-trapped closed $2$-surface one can use to foliate trapped and anti-trapped regions.
\item An untrapped region is a region which can be foliated by a choice of untrapped closed $2$-surfaces, as in standard Minkowski spacetime. Within that region admitting such foliation, the expansion of the ingoing and outgoing light rays are characterized by
\begin{equation}
\theta_{+} > 0, \quad \theta_{-} < 0, \qquad \text{and} \qquad \theta_{+} \theta_{-} < 0,
\end{equation}
which implies that outgoing light wave-fronts are expanding while ingoing ones are contracting, as expected in a normal region. In this untrapped region, $H_{\perp}$ is timelike while $H$ is spacelike.
\item A trapped region corresponds to a region which can be foliated by a choice of trapped closed $2$-surfaces such that in that region, the expansions of in-going and out-going light rays satisfy
\begin{equation}
\theta_{+} < 0, \quad \theta_{-} < 0, \qquad \text{and} \qquad \theta_{+} \theta_{-} > 0.
\end{equation}
In a trapped region, $H_{\perp}$ spacelike. Conversely, $H$ is timelike inside.
\item  An anti-trapped region which can foliated by a choice of anti-trapped closed $2$-surfaces such that within that region, the expansions of light rays behave as
\begin{equation}
\theta_{+} > 0, \quad \theta_{-} > 0, \qquad \text{and} \qquad \theta_{+} \theta_{-} > 0.
\end{equation}
In an anti-trapped region, one obtains the same. Therefore, the sign of the norm of the MCV and its dual can only distinguish between untrapped versus (anti-)trapped regions, but not between antitrapped and trapped ones.  
\end{itemize}
One can further characterize the nature of the horizon by studying how the expansions change along each null directions. Indeed, as initially introduced by Hayward in \cite{Hayward:1993wb}, the sign of the Lie derivative of the expansion along one of the null directions encodes the outer/inner nature of the (anti)-trapping horizon and allows one to discriminate between the horizon of compact objects (black and white holes) and cosmological horizons. We distinguish the four different cases:
\begin{itemize}
\item A future-outer horizon (corresponding to a black hole) satisfies
\be
\cL_{-} \theta_{+} < 0,
\ee
which means that crossing the horizon in the inward null direction, one moves from an untrapped to a trapped region.
\item A future-inner horizon (corresponding to a contracting cosmological horizon) satisfies
\be
\cL_{-} \theta_{+} > 0,
\ee
meaning that crossing this horizon in the inward null direction $\ell_{-}$, one moves from an trapped to a untrapped region. An example is provided by the inner (contracting) cosmological horizon of closed FLRW universe used to model the interior of the collapsing star in the Oppenheimer-Snyder model.
\item A past-outer horizon (corresponding to white hole) satisfies 
\be
\cL_{+} \theta_{-} < 0,
\ee
which encodes the fact that moving along the outward null direction, one moves from an anti-trapped to an untrapped region. 
\item Finally, a past-inner horizon (corresponding to an expanding cosmological horizon) satisfies 
\be
\cL_{+} \theta_{-} > 0,
\ee
In this case, moving along the outward null direction, one moves from an untrapped to an anti-trapped region.  
\end{itemize}

Therefore, one can compute the norm of the mean curvature vector (or its dual) and track the presence of (anti-)trapping horizons by identifying the locus of points where its norm vanishes. This provides a concrete extension of the well-known properties of the Kodama vector, as the utilization of the MCV is not confined to spherical symmetry. This was already noticed in \cite{Senovilla:2014ika}. See \cite{Dotti:2023elh, Dotti:2025npw} for related work using the MCV to locate horizons. However, as we shall see, the use of the MCV still involves a degree of ambiguity related to the embedding of the surface $\mathcal{S}$. This issue becomes apparent in certain axisymmetric geometries, and it is absent only in spherical symmetry or under the assumption of stationarity.

\subsubsection{Relation to the Kodama vector}

Let us relate the Kodama vector to the dual mean curvature vector introduced above. The induced metric $q$ on the closed surface  $\cS$ is\footnote{Notice that, at this stage, the embedding of $\mathcal{S}$ has already been fixed. We consider a two-surface defined by constant $t$ and $r$. This choice is consistent with the assumption of spherical symmetry, adopted to connect with the Kodama construction, where the horizon is necessarily characterised by $r = \mathrm{const}$.}
\be
\rd s^2_{\cS} = q_{\mu\nu} \rd x^{\mu} \rd x^{\nu} = R^2(t,r) \rd \Omega^2,
\ee
while the two unit normal vectors $(n,s)$ are given by
\begin{align}
& n_{\mu} \rd x^{\mu} = - e^{-\Phi} \sqrt{f} \rd t, \qquad s_{\mu} \rd x^{\mu} = \frac{1}{\sqrt{f}} \rd r, \\
& \;\; n^{\mu} \partial_{\mu} = \frac{e^{\Phi}}{\sqrt{f}} \partial_t, \qquad \;\;\; \;\; s^{\mu} \partial_{\mu} = \sqrt{f} \partial_r,
\end{align}
of which the corresponding extrinsic curvatures read
\begin{align}
 & K_{\theta\theta} (n) =  \frac{K_{\varphi\varphi} (n)}{\sin^2{\theta}} = \frac{e^{\Phi} R \dot{R}}{\sqrt{f}}, \qquad K_{\theta\theta} (s) =  \frac{K_{\varphi\varphi} (s)}{\sin^2{\theta}} = - \sqrt{f} R R', \\
 & K(n) = \frac{2 e^{\Phi}}{\sqrt{f}} \frac{\dot{R}}{R}, \qquad \qquad \qquad \qquad \;\;\; K(s) = - 2 \sqrt{f} \frac{R'}{R} .
\end{align} 
The mean curvature vector and its dual are therefore given by
\begin{align}
H^{\mu} \partial_{\mu} & = - \frac{1}{R} \left[ f R' \partial_r + \frac{e^{2\Phi}}{f} \dot{R} \partial_t\right],\\
H_{\perp}^{\mu} \partial_{\mu} & = - \frac{ e^{\Phi}}{R} \left[ \dot{R} \partial_r +  R' \partial_t\right],
\end{align}
results that show how the dual vector $H_{\perp}$ is proportional to the Kodama vector, i.e.,
\begin{equation}
H^{\mu}_{\perp} \partial{\mu} = - R^{-1} K^{\mu} \partial_{\mu}.
\end{equation}
One can easily verify that
\be
\label{nor}
g_{\mu\nu} H^{\mu} H^{\nu} = -  g_{\mu\nu} H^{\mu}_{\perp} H^{\nu}_{\perp} = 
{- \frac{ e^{2\Phi} }{R^2} \left[ \frac{\dot{R}^2}{f} - e^{-2\Phi} f (R')^2\right] } =- \frac{ e^{2\Phi}}{R^2}  K_{\mu} K^{\mu}.
\ee
This clarifies the relation between the objects introduced in the general case, where spherical symmetry is not assumed, and the special case of spherical symmetry, where the Kodama vector is defined. In particular, using the property (\ref{exp}), one recovers that the Kodama vector is null on any (anti-)trapping horizon.  To see this, let us apply this construction to two well-known examples. First, consider the Schwarzschild black hole geometry for which the form of the metric (\ref{metSS}) implies that $\Phi(t,r) = 0$, $R(t,r) = r$, and $m(t,r) = m$. Then, the norms of the mean curvature and Kodama vectors are given by
\begin{equation}
g_{\mu\nu} H_{\perp}^{\mu} H_{\perp}^{\nu} = \frac{e^{2\Phi}}{R^2} K^{\mu} K_{\mu} = - \frac{(r - 2m)}{r^3}.
\end{equation}
As expected, the dual mean curvature vector and the Kodama vector are timelike for $r > 2m$, null at $r = 2m$, i.e., at the Schwarzschild horizon, and spacelike for $r < 2m$. 

 Second, let us consider the flat FLRW geometry. In the form (\ref{metSS}), the FLRW metric corresponds to the functions $e^{-2\Phi(t)}=a^4$, $f(t) =1/a^2(t)$ and $R(t,r) = a(t)r$ where $a(t)$ is the scale factor and where the time corresponds to the conformal time. Applying the formula (\ref{nor}), one obtains
\be
g_{\mu\nu} H_{\perp}^{\mu} H_{\perp}^{\nu}  = -\frac{1}{a^2(t) r^2} \left( \frac{r^2 \dot{a} ^2}{a^2} -1\right)
\ee
As expected, one recovers that the FLRW horizon corresponds to the hypersurface $r = H^{-1}(t)$ where $H =\dot{a} ^2 / a^2$ is the Hubble factor. In the untrapped region corresponding to the interior region, i.e. $r < H^{-1}(t)$, the MCV is indeed timelike while beyond the cosmological horizon, i.e. $r > H^{-1}(t)$, it is spacelike. Finally, it becomes null on the cosmological horizon.

This concludes our review on the generalization of the Kodama vector beyond spherical symmetry and on the foliation-independent mechanism to identify trapping horizons in axisymmetric dynamical black hole configurations.
In the next section, we focus on the second objective of this work, namely the development of a solution-generating technique capable of producing dynamical axisymmetric exact solutions.


\section{Solution generating techniques for axisymmetric primordial black holes}

\label{SEC2}
In this section, we present a new solution-generating technique to construct exact dynamical and axisymmetric solutions of the Einstein–Scalar system with a self-interacting potential. As we shall see, these solutions describe asymptotically FLRW axisymmetric black (and white) holes. We first review the existing solution-generating methods for constructing (i) static and axisymmetric or (ii) spherically symmetric dynamical black hole solutions in the Einstein–Scalar system. We then show how these two methods can be combined to build dynamical axisymmetric black hole solutions.

\subsection{Building axisymmetric or time-dependent black holes}\label{FONAREV}

A rather natural starting point to obtain exact solutions describing primordial black holes is to consider GR sourced by either a perfect fluid or a self-interacting scalar field. In the following, we shall focus on the latter, as in this case it is possible to consider a well-motivated action principle from which the field equations can be deducted. Within this context, the no-hair theorem already implies that none of the stationary solutions of such systems can describe a black hole (they usually correspond to naked singularities). However, relaxing stationarity allows one to evade the no-hair theorem such that dynamical trapped regions can form. 

In the context of Einstein-scalar theory, a good example of this is provided by the well-known Fischer-Janis-Newman-Winicour (FJNW) solution obtained in \cite{Janis:1968zz}, which describes a static and asymptotically flat naked singularity. As first shown in \cite{Husain:1994uj} by Husain, Nunez, and Martinez (HNM), this solution can be generalised to a dynamical, asymptotically FLRW geometry, which now contains  trapped (or anti-trapped) regions. 

For spherically symmetric scalar vacuum geometries\footnote{Here, scalar-vacuum solutions are understood as the set of exact solutions of the Einstein-scalar system, including the presence of a self-interacting potential.}, the relaxation of stationarity has been systematically studied by Fonarev in \cite{Fonarev:1994xq}, providing a powerful solution-generating method to construct time-dependent scalar vacuum solutions starting from static ones. Yet, so far, this method has been restricted to the consideration of static and spherically symmetric seed geometries. 

A natural extension of these results consists of relaxing the assumptions of spherical symmetry and staticity on the given seed. This leads, in increasing order of complexity, to the study of the static axisymmetric case and its stationary generalisation. To pursue this programme, it is first necessary to establish a systematic procedure for coupling a given vacuum spacetime to a minimally coupled scalar field. In this regard, any static and axisymmetric vacuum solution, regardless of the presence of additional symmetries, can be promoted to a solution of the Einstein–Scalar theory through Buchdahl’s theorem \cite{Buchdahl:1959nk,Barrientos:2024uuq}. As we shall see, this result naturally combines with Fonarev’s method, enabling the construction of dynamical hairy spacetimes beyond spherical symmetry.
Relaxing staticity, however, is a more delicate task. Buchdahl’s theorem applies only under the assumption of staticity, at least in four dimensions. However, the theorem of Eris and Gürses, which is valid for any circular, stationary, and axisymmetric spacetime—i.e., for the entire Weyl–Lewis–Papapetrou class—offers a broader mechanism to construct exact solutions within the Einstein–Scalar theory in the stationary and axisymmetric setting \cite{Barrientos:2025abs}. Whether the general framework of Eris–Gürses can be consistently adapted to Fonarev’s scheme to generate dynamical rotating configurations remains an open question. In the appendix \ref{appA}, we discuss the main obstacles that prevent a straightforward combination of the two approaches.

\subsubsection{From spherical symmetry to axisymmetry: Buchdahl method}

The original Buchdahl method was constructed within the context of the Einstein-scalar system. Consider, therefore, the following action
\be
\label{EMm}
\cS = \int d^4x \sqrt{|g|} \left( \frac{\cR}{2\kappa} - \frac{1}{2} g^{\mu\nu} \phi_{\mu} \phi_{\nu} \right)  ,
\ee
where $\phi$ is a massless scalar field and $\phi_\mu$ is a shorthand notation for $\nabla_\mu\phi$. Here, $\kappa = 8\pi G$ and the dimensions are given by $[\cR] = L^{-2}$, $[\phi] = L^{-1}$. The field equations read
\begin{align}
\label{efe1}
G_{\mu\nu} & = \kappa T_{\mu\nu} = \kappa \left( \phi_{\mu} \phi_{\nu} - \frac{1}{2}g_{\mu\nu}   g^{\alpha\beta} \phi_{\alpha} \phi_{\beta} \right),\\
\label{efe2}
\Box \phi & = 0.
\end{align}
In general, solving these field equations via brute force turns out to be difficult. The simplest solution, which is spherically symmetric and static, was found in \cite{Fisher:1948yn, Janis:1968zz}
\begin{align}
\label{FJNWsol}
    ds_{FJNW}^2&=-\left(1-\frac{2m}{r}\right)^\beta dt^2+\frac{dr^2}{\left(1-\frac{2m}{r}\right) ^\beta}+\left(1-\frac{2m}{r}\right)^{1-\beta} r^2 d\Omega^2,\\
    \phi&=\sqrt{\frac{1- \beta^2}{2\kappa}}\ln\left(1-\frac{2m}{r}\right),
\end{align}
and it is known as the FJNW configuration.
It is easy to see that it reduces to the Schwarzschild black hole if the so-called hair parameter $\beta$ goes to $1$. However, the nature of this solution is very different from the Schwarzschild one, as it describes a one parameter family of naked singularities.  Notice that, contrary to the Schwarzschild singularity, where the central singularity corresponds to a spacelike hypersurface located at $r=0$ (recall that  the radial and time directions are switched when crossing the horizon such that $r$ is a time coordinate in the black hole interior, i.e. for $r<2m$), the FJNW solution features a singularity on the timelike hypersurface $r= 2m$. Moreover, one can show that no regular horizon can exist if $\beta\neq 1$, i.e., when the geometry deviates from the Schwarzschild black hole, which can be understood as a consequence of the no-hair theorem. 

Interestingly, provided one knows a static and therefore spherically symmetric vacuum solution (without a scalar field), one can construct a static and axisymmetric scalar vacuum solution, thanks to the Buchdahl method \cite{Buchdahl:1956zz, Buchdahl:1959nk}. Notice that this method was originally used onto spherically symmetric seeds; however, the application to axisymmetric ones was understood to be straightforward \cite{Barrientos:2024uuq}. The Buchdahl method works as follows. Consider a vacuum solution $\bar{g}$ with line element 
\begin{equation}
    d\bar{s}^2= \bar{g}_{\mu\nu} d x^{\mu} d x^{\nu} = \bar{g}_{00}d\eta^2+\bar{h}_{ij}dx^idx^j,\quad  with \quad x^\mu=(x^0,x^i)=(\eta,x^i),\quad i=1,2,3 \label{buchdahlmetric}
\end{equation}
with no summation understood in "$x^0=\eta$", and where the $\eta-$coordinate satisfies 
\begin{equation}
    \partial_0 \bar{g}_{\mu\nu}=0= \bar{g}_{i0}, \label{condbuch}
\end{equation}
namely, it is an ignorable coordinate along which the spacetime has no off-diagonal metric components \footnote{For clarity, we use the subscript-0 for components along the $\eta$ direction (e.g., $\bar{g}_{00}$). When allowing explicit $\eta$-dependence, however, we instead write it directly, as in the conformal factor $\mu(\eta)$ (see below).}. A solution $(g, \phi)$ in the Einstein-scalar theory (\ref{EMm}) will then be given by 
    \begin{align}
    \label{sol1}
    ds^2_{ES}&=g_{\mu\nu}dx^\mu dx^\nu=( \bar{g}_{00})^\beta d\eta^2+(\bar{g}_{00})^{1-\beta} \bar{h}_{ij}dx^idx^j,\\
        \label{sol2}
    \phi&=\xi_0\ln(\bar{g}_{00}),
\end{align}
where $\xi_0$ is related to the scalar charge $\beta$ through
\be
\xi_0 = \sqrt{\frac{1-\beta^2}{2\kappa}}.
\ee

An intuitive reading of condition \eqref{condbuch} suggests identifying $\eta$ with the time coordinate, in accordance with the requirements for staticity. Proceeding in this way, one immediately recovers the FJNW configuration \eqref{FJNWsol} from the vacuum Schwarzschild solution. However, as pointed out in \cite{Barrientos:2024uuq}, the applicability of Buchdahl’s theorem is not confined to spherically symmetric seeds, nor does it require interpreting $\eta$ as the temporal coordinate. In fact, the theorem can be employed more generally by treating 
$\eta$ as any ignorable coordinate of the seed metric, independently of its relation to staticity.
This observation opens two distinct avenues. First, if the seed is spherically symmetric, axisymmetry can be introduced by assigning $\eta$ to the azimuthal coordinate. This approach has been extensively explored in \cite{Barrientos:2024uuq}, leading to the characterisation of a broad class of vacuum spacetimes with asymptotically Levi-Civita behavior. In addition, a rotating extension of this family has recently been constructed \cite{Barrientos:2025rjn} by exploiting the relation between Buchdahl’s theorem and the discrete inversion symmetry inherent in the Einstein equations when expressed in terms of the Ernst complex potential formalism \cite{Ernst:1967wx,Ernst:1967by}. 
Second, one may start directly from an axisymmetric seed, where a concrete example is given by the Einstein–Scalar extension of the vacuum Zipoy–Voorhees spacetime, recently constructed in \cite{Barrientos:2025abs, Azizallahi:2023rrv}.
In particular, and of greater relevance to our purposes, the Buchdahl framework enables the construction of axisymmetric solutions with a scalar field, where the scalar profile is necessarily proportional to the metric function associated with the differential of the coordinate $\eta$ on which the theorem is applied. As we will see, this feature plays a crucial role in combining the Buchdahl and Fonarev techniques, thereby providing a unified framework for constructing dynamical, axisymmetric, hairy black hole configurations.

We now turn to the second key theorem, introduced by Fonarev, which provides a mechanism to relax the assumption of stationarity.

\subsubsection{From static to dynamical geometries: Fonarev method}

As mentioned earlier, the derivation of exact time-dependent black hole solutions in GR is a challenging task. A first example was provided by the HMN scalar vacuum solution, which corresponds to a time-dependent generalisation of the static FJNW solution (\ref{FJNWsol}) describing a scalar collapse\footnote{See also \cite{Roberts:1989sk}.} \cite{Husain:1994uj}. Around the same time, Fonarev provided a systematic understanding of the solution-generating map, allowing for such time-dependent inhomogeneous collapsing/expanding scalar vacuum solution \cite{Fonarev:1994xq}. The framework of this method is the Einstein-scalar system with a self-interacting Liouville potential $V({\tilde{\phi}})$ whose action reads
\be
\cS = \int d^4x \sqrt{|\tilde{g}|} \left( \frac{\tilde{\cR}}{2\kappa} - \frac{1}{2} g^{\mu\nu} \tilde{\phi}_{\mu} \tilde{\phi}_{\nu} - V(\tilde{\phi})\right),
\ee
where $V(\tilde{\phi}) = V_{0} e^{\xi_3 \tilde{\phi}}$, and units are such that $[V_0] = L^{-4}$ and  $[\xi_3] = [L]$. The field equations are given by
\begin{align}
\label{efe11}
\tilde{G}_{\mu\nu} & = \kappa \tilde{T}_{\mu\nu} = \kappa \left[ \tilde{\phi}_{\mu} \tilde{\phi}_{\nu} - \tilde{g}_{\mu\nu} \left( \frac{1}{2} \tilde{g}^{\alpha\beta} \tilde{\phi}_{\alpha} \tilde{\phi}_{\beta} + V(\tilde{\phi}) \right)\right],\\
\label{efe22}
\tilde{\Box} \tilde{\phi} & = V_{\tilde{\phi}}.
\end{align}
The Fonarev solution generating map can be stated as follows. 

Consider an exact static and spherically symmetric solution $(g, \phi)$ of the field equations (\ref{efe1} - \ref{efe2}) of the form (\ref{sol1} - \ref{sol2}). Then, a time-dependent extension $(\tilde{g}, \tilde{\phi})$, which now solves the field equations with a Liouville potential $V=V_0e^{\xi_3\tilde{\phi}}$, is given by
\begin{align}
    d\tilde{s}^2&=e^{2\mu(t)}d s^2,\\
    \tilde{\phi} &=\phi +\frac{\xi_1}{\kappa}\mu(t),
\end{align}
where the conformal factor is 
\be
\mu(t)=\xi_2\ln(Ct+B),
\ee
being $C$ and $B$ two free constants while the parameters $V_0$, $\xi_1$, $\xi_2$, and $\xi_3$ are constrained to follow 
\begin{align}
    &\xi_{1}= - \xi_3 = \frac{\beta}{\xi_{0}}, \\
    & (\beta^{2}-2\xi_{0}^{2}\kappa)\xi_{2}=2\xi_{0}^{2}\kappa,  \\
    & (2\xi_{0}^{2}\kappa-\beta^{2})^{2} V_{0}=- \; 2\xi_{0}^{2}C^{2}(\beta^{2}-6\xi_{0}^{2}\kappa).\label{coefff}
\end{align}
This provides a general technique to obtain exact solutions describing spherically symmetric scalar collapse embedded in a FLRW universe. Notice that $B$ can always be absorbed by a redefinition of the time, so that we shall not consider it in the following. 
 
For instance, consider the FJNW static configuration (\ref{FJNWsol}). Applying the Fonarev theorem gives the new exact time-dependent solution
\begin{subequations}
\label{FJNW}
\begin{align}
    d\tilde{s}^2&= (Ct)^{\xi_2} \left[ -\left(1-\frac{2m}{r}\right)^\beta dt^2+\frac{dr^2}{\left(1-\frac{2m}{r}\right) ^\beta}+\left(1-\frac{2m}{r}\right)^{1-\beta} r^2 d\Omega^2 \right] ,\\
    \tilde{\phi}&=\xi_0 \ln\left(1-\frac{2m}{r}\right) + \frac{\xi_1 \xi_2}{\kappa}\ln(Ct), 
\end{align}
\end{subequations}
where the parameter space defined by $V_0$, $\xi_1$, $\xi_2$ and $\xi_3$ is constrained as stated above.  The HMN scalar collapse solution derived in \cite{Husain:1994uj} for the massless system is obtained by setting $V_0 =0$ and $\beta^{2} = 6\xi_{0}^{2} \kappa$, which is solved for the specific value $\beta = \pm \sqrt{3}/2$. For $\beta \neq \sqrt{3}/2$, the family of solutions obtained through this method corresponds to those derived by Fonarev in \cite{Fonarev:1994xq}. To our knowledge, these provide the only known examples of exact inhomogeneous scalar collapse solutions of the self-interacting Einstein-scalar system in the literature.

Now, although the theorem has been initially tailored to act on static and spherically symmetric configurations and to add time dependence, it can be easily generalised in the following two directions: 
\begin{itemize}
\item (i) spherical symmetry is not essential, as under some conditions, axisymmetric line elements can also be accommodated, and
\item  (ii) the initial configuration can be upgraded to depend on any of the ignorable coordinates of the seed and not exclusively on the time coordinate. 
\end{itemize}
Both generalisations can be achieved by considering a general static and axisymmetric line element and by performing a superposition of the Fonarev and Buchdahl theorems, which we now describe.

\subsection{Building axisymmetric time-dependent solutions}

In this section, we discuss how one can construct exact analytic solutions of GR that describe dynamical and axisymmetric compact objects embedded in an FLRW cosmology.  Focusing on the Einstein-scalar system, it is natural to wonder whether the two complementary methods reviewed above can be combined to construct more realistic dynamical and axisymmetric solutions beyond the spherically symmetric sector. In the following, we present a generalisation that allows one, given a stationary axisymmetric scalar vacuum solution, to systematically construct its time-dependent extension.  

\subsubsection{Extended Fonarev theorem}
\begin{theorem}
\label{extendedfonarev} Consider now the static and axisymmetric vacuum solution $\bar{g}$\footnote{For our purposes, we require that the full metric \eqref{buchdahlaxisymmetric} be axisymmetric.}  
\begin{align}
    d\bar{s}^2= \bar{g}_{\mu\nu} \rd x^{\mu} \rd x^{\nu} = \bar{g}_{00}d\eta^2+\bar{h}_{ij}dx^idx^j, \label{buchdahlaxisymmetric}
\end{align}
 and that satisfies
 \begin{equation}
    \partial_0 \bar{g}_{\mu\nu}=0= \bar{g}_{i0}, 
\end{equation}
i.e. no $x^0=\eta-$dependence. Then, one can construct an $\eta$-dependent extension $(\tilde{g}, \tilde{\phi})$ that solves the Einstein–Scalar system with the self-interacting potential
$V(\tilde{\phi})=V_0e^{\xi_3\tilde{\phi}}$, and which takes the form
\begin{align}
    d \tilde{s}^2&=e^{2\mu(\eta)}[(\bar{g}_{00})^\beta d\eta^2+(\bar{g}_{00})^{1-\beta} \bar{h}_{ij}dx^idx^j],\\
    \tilde{\phi}&=\xi_0\ln(\bar{g}_{00})+\frac{\xi_1}{\kappa}\mu(\eta),
\end{align}
with conformal factor
\be
\mu(\eta)=\xi_2\ln(C\eta+B).
\ee
The parameter space defined by $V_0$, $\xi_1$, $\xi_2$, and $\xi_3$ is constrained to follow 
\begin{subequations}
\label{coeff}
\begin{align}
    &\xi_{1}= - \xi_3 = \frac{\beta}{\xi_{0}} \\
    & (\beta^{2}-2\xi_{0}^{2}\kappa)\xi_{2}=2\xi_{0}^{2}\kappa  \\
    & (2\xi_{0}^{2}\kappa-\beta^{2})^{2} V_{0}=\mp \; 2\xi_{0}^{2}C^{2}(\beta^{2}-6\xi_{0}^{2}\kappa),
\end{align}
\end{subequations}
with $(-)$ if $\eta$ is timelike and $(+)$ if spacelike, and with constants $C$ and $B$ remaining free. Once more, the parameter $B$ can be removed by a redefinition of the coordinate $\eta$, so that we shall omit it in the following. 
\end{theorem}

Next, we present the full development of the proof, followed by an explicit example of an axisymmetric, time-dependent black hole solution of the Einstein–Scalar system.

\subsubsection{Proof of extended Fonarev's theorem \label{A}}

To prove the extended Fonarev's theorem (\ref{extendedfonarev}) we start from the following axially symmetric vacuum metric $\bar{g}$ which reads
\begin{equation}
    d\bar{s}^2= \bar{g}_{\mu\nu} \rd x^{\mu} \rd x^{\nu} = \bar{g}_{00}d\eta^2+\bar{h}_{ij}dx^idx^j,  \label{buchdahlmetricproof}
\end{equation}
again, with no summation understood in "$\eta$", and where the $\eta-$coordinate satisfies  
\begin{equation}
    \partial_0 \bar{g}_{\mu\nu}=0= \bar{g}_{i0}.  
\end{equation}
According to Buchdahl's theorem \cite{Buchdahl:1959nk}, from \eqref{buchdahlmetricproof} we can construct the following solution $(g,\phi)$ of the Einstein-scalar system without self-interaction which reads
\begin{subequations}
   \label{confproof}
    \begin{align}
    ds^2_{ES}&= g_{\mu\nu} \rd x^{\mu} \rd x^{\nu} = (\bar{g}_{00})^\beta(dx^a)^2+(\bar{g}_{00})^{1-\beta} \bar{h}_{ij}dx^idx^j,\label{ESansatz}\\
\phi&=\xi_0\ln(\bar{g}_{00})\label{ESscalar},
\end{align}
\end{subequations}
where we recall that, for simplicity, we have defined 
\be
\xi_0=\sqrt{\frac{1-\beta^2}{2\kappa}} \qquad \text{with} \qquad  -1 \leqslant \beta \leqslant 1.
\ee
In order to construct the extended Fonarev's theorem we consider the following enhanced configuration
\begin{align}
    d \tilde{s}_{}^2&= \tilde{g}_{\mu\nu} \rd x^{\mu} \rd x^{\nu} = e^{2\mu(\eta)}ds^2_{ES},\\
    \tilde{\phi}&=\xi_0\ln(\bar{g}_{00})+\xi_1\Psi(\eta)=\phi+\xi_1\Psi(\eta),
\end{align}
in which the Einstein-scalar line element \eqref{ESansatz} has been multiplied by a conformal factor depending on the ignorable coordinate and the scalar profile \eqref{ESscalar} has been extended by a function of the same coordinate. As stated by the theorem, this configuration will now solve the Einstein equations for a self-interacting scalar field with a Liouville potential given by (\ref{efe11} - \ref{efe22}). 

Hence, to proceed with the proof, we first compute the conformally transformed Ricci tensor $\tilde{R}_{\mu\nu}$, which yields 
\begin{align}
    \tilde{R}_{\mu\nu}&=R_{\mu\nu}-g_{\mu\nu}\Box\mu-2[\nabla_{\mu}\nabla_{\nu}\mu+g_{\mu\nu}\nabla_{\gamma}\mu\nabla^{\gamma}\mu-\nabla_{\mu}\mu\nabla_{\nu}\mu]. \label{riccitensorfonarev}
\end{align}
Each of the differential operators, once evaluated on $\mu=\mu(\eta)$, recalling $\eta$ as an ignorable coordinate, reduces to  
\begin{align}
    \Box\mu&=(\bar{g}^{00})^{-\beta}\Ddot{\mu}-g^{\alpha\beta}\Gamma^{0}_{\alpha\beta}\Dot{\mu},\\
    \nabla_{\alpha}\nabla_{\beta}\mu&=\Ddot{\mu}-\Gamma^{0}_{\alpha\beta}\Dot{\mu},\\
    \nabla_{\alpha}\mu\nabla^{\alpha}\mu&=(\bar{g}^{00})^{-\beta}\Dot{\mu}^{2},
\end{align}
with the dot symbol denoting differentiation with respect to $\eta$. For the line element \eqref{ESansatz}, the only non-vanishing component of the  Christoffel connection is 
\begin{align}
    {\Gamma}^{0}_{0k}&=\frac{1}{2}(\bar{g}^{00})^{-\beta}\partial_{k}(\bar{g}_{00})^{\beta},
\end{align}
where $k$ collectively denotes all non-Killing coordinates. With these considerations at hand, the Ricci tensor \eqref{riccitensorfonarev} reads 
\begin{align}
    \tilde{R}_{\mu\nu}&=R_{\mu\nu}-g_{\mu\nu}(\bar{g}^{00})^{-\beta}\Ddot{\mu}-2[\partial_{\mu}\partial_{\nu}\mu-\Gamma^{0}_{\mu\nu}\Dot{\mu}+g_{\mu\nu}(\bar{g}^{00})^{-\beta}\Dot{\mu}^{2}-\partial_{\mu}\mu\partial_{\nu}\mu], 
\end{align}
of which the non-vanishing components are 
\begin{align}
    \tilde{R}_{00}&=R_{00}-3\Ddot{\mu},\\
    \tilde{R}_{0k}&=R_{0k}+(\bar{g}^{00})^{-\beta}\partial_{k}(\bar{g}_{00})^{\beta}\Dot{\mu},\\
    \tilde{R}_{kl}&=R_{kl}-g_{kl}(\bar{g}_{00})^{1-\beta}(\bar{g}^{00})^{-\beta}(\Ddot{\mu}+2\Dot{\mu}^{2}).
\end{align}
At this point, it is convenient to work with Einstein-scalar equations written in the alternative fashion $\tilde{R}_{\mu\nu}=\kappa \tilde{S}_{\mu\nu}$, where $\tilde{S}_{\mu\nu}=\partial_{\mu}\tilde{\phi}\partial_{\nu}\tilde{\phi}+\tilde{g}_{\mu\nu}V(\tilde{\phi})$.
Having evaluated the Ricci tensor, we now proceed with the evaluation of the energy-momentum tensor. Its components yield
\begin{align}
    \tilde{S}_{00}&=S_{00}+\xi_{1}^{2}\Dot{\Psi}^{2}+(\bar{g}_{00})^{\beta}e^{2\mu}V,\\
    \tilde{S}_{0k}&=S_{0k}+\xi_{1}\partial_{k}\phi\Dot{\Psi},\\
    \tilde{S}_{kl}&=S_{kl}+g_{kl}(\bar{g}_{00})^{1-\beta}e^{2\mu}V, 
\end{align}
where $S_{\mu\nu}=\partial_\mu\phi\partial_\nu\phi$ relates to the energy-momentum tensor of configuration \eqref{confproof}, the initial static seed. Finally, we are left with the system of equations 
\begin{subequations}
\begin{align}
    -3\Ddot{\mu}&=\kappa(\xi_{1}^{2}\Dot{\Psi}^{2}+(\bar{g}_{00})^{\beta}e^{2\mu}V),\label{firsteinsteinproof}\\
    (\bar{g}^{00})^{-\beta}\partial_{k}(\bar{g}_{00})^{\beta}\Dot{\mu}&=\kappa\xi_{1}\partial_{k}\phi\Dot{\Psi}\label{secondeinsteinproof},\\
    -(\bar{g}^{00})^{-\beta}(\Ddot{\mu}+2\Dot{\mu}^{2})&=\kappa e^{2\mu}V. \label{thirdeinsteinproof}
\end{align}
\end{subequations}
From equation \eqref{secondeinsteinproof} the following condition becomes necessary 
\begin{align}
    (\bar{g}^{00})^{-\beta}\partial_{k}(\bar{g}_{00})^{\beta}\propto\partial_{k}{\phi}.
\end{align}
Accordingly, we consider $\partial_k\phi=P(\bar{g}^{00})^{-\beta}\partial_{k}(\bar{g}_{00})^{\beta}$, with $P$ a proportionality constant. Here it relies the utility of Buchdahl theorem \cite{Buchdahl:1959nk}, as it always provides us with a seed solution that satisfies this condition and hence equation \eqref{secondeinsteinproof} transforms into 
\begin{align}
\Dot{\mu}&=\kappa\xi_{0}\xi_{1}P\Dot{\Psi},
\end{align}
of which the solution, using $\xi_{0}\xi_{1}P=1$, is 
\begin{align}
    \Psi(\eta)&=\frac{\mu(\eta)}{\kappa}.
\end{align}
Acting on this result for \eqref{firsteinsteinproof} and \eqref{thirdeinsteinproof}, and combining them properly, provide the relations 
\begin{subequations}
\begin{align}
    \Ddot{\mu}+\left( \frac{\xi^{2}_{1}-2\kappa}{2\kappa} \right) \Dot{\mu}^{2}&=0,\label{firstdiffequation}\\
\Dot{\mu}^{2}\left(\frac{\xi_{1}^{2}-6\kappa}{\kappa}\right)-2(\bar{g}_{00})^{\beta}\kappa^{2}e^{2\mu}V&=0.\label{seconddiffequation}
\end{align}
\end{subequations}
Equation \eqref{firstdiffequation} delivers the conformal factor 
\begin{equation}
    \mu(\eta)=\xi_2 \ln (C\eta),
\end{equation}
with $\xi_2$ to be fixed, $C$, and a  free constant. Replacing the conformal factor in \eqref{seconddiffequation}, and taking the potential to be $V(\Psi)=V_0e^{\xi_3\tilde{\phi}}$, finally solves all field equations provided the following constraints between the parameters hold
\begin{align}
    &\xi_{0}=\sqrt{\frac{1-\beta^{2}}{2\kappa}},\hspace{0.2cm}\xi_{1}=\frac{\beta}{\xi_{0}},\hspace{0.2cm}\xi_{2}=\frac{2\xi_{0}^{2}\kappa}{\beta^{2}-2\xi_{0}^{2}\kappa},\hspace{0.2cm}\xi_{3}=-\frac{\beta}{\xi_{0}},\hspace{0.2cm}
    V_{0}=(\mp)\frac{2\xi_{0}^{2}C^{2}(\beta^{2}-6\xi_{0}^{2}\kappa)}{(2\xi_{0}^{2}\kappa-\beta^{2})^{2}}.
\end{align}
Notice that, by construction, the solution with a non-vanishing potential is defined for $2\xi_{0}^{2}\kappa-\beta^{2} \neq 0$, which implies, $\beta \neq \pm \sqrt{\frac{2}{3}}$.
The different parameters $(\xi_1, \xi_2, \xi_3)$ can be expressed in term of $\beta$ as 
\be
\xi_1 = - \xi_3 = \sqrt{\frac{2\kappa \beta}{1 -\beta^2}}, \qquad \xi_2 = - \frac{1-\beta^2}{1-2\beta^2}.
\ee
Hence, explicitly, the final $\eta$-dependent configuration reads 
\begin{align}
    ds^{2}&=(C\eta)^{2\xi_{2}}[(g_{00})^{\beta}d\eta^{2}+(g_{00})^{1-\beta}g_{ij}dx^{i}dx^{j}],\\
\Phi&=\xi_{0}\ln(g_{00})+\frac{\xi_{1}\xi_{2}}{\kappa}\ln(C\eta).
\end{align}
Notice that we recover the HMN exact solution \cite{Husain:1994uj} of the massless Einstein-scalar system by considering $\eta$ as the time coordinate, and for the values 
$\beta^{2}-6\xi_{0}^{2}\kappa = 0$, for which $V_0 =0$ and  $\beta = \pm \frac{\sqrt{3}}{2}$.

This concludes the presentation and proof of the extended solution–generation technique, which enables the construction of exact dynamical axisymmetric solutions of the Einstein–Scalar system. We are now in a position to apply this method and present a concrete example of an axisymmetric, asymptotically FLRW, time-dependent black hole.

\section{The Zipoy-Voorhees time-dependent solution}

\label{SEC3}

\label{example}

In this section, we present a first explicit example of an exact, asymptotically FLRW, axisymmetric solution of the Einstein–Scalar system. We begin by describing the seed geometry. 

\subsection{The static seed}

As stated above, the Fonarev scheme was originally developed for the construction of dynamical spacetimes with spherical symmetry. We have shown, however, that this assumption can be relaxed to encompass spacetimes with axisymmetry only. By virtue of the extended Fonarev theorem (\ref{extendedfonarev}) established in the previous section, we have demonstrated that dynamical geometries can be generated directly from an axisymmetric static seed. 

As is well known, in four dimensions the Weyl problem, namely the task of solving the vacuum Einstein field equations for the most general static and axisymmetric line element, has been fully resolved from a mathematical standpoint. As a result, an infinite family of static, axisymmetric spacetimes can be constructed analytically, with the axisymmetric sector of the metric expressed through a complete multipolar expansion.
Within this family, the Zipoy–Voorhees (ZV) spacetime \cite{Zipoy:1966btu,Voorhees:1970ywo} constitutes an axisymmetric generalisation of the Schwarzschild geometry that incorporates asymptotically flat multipolar corrections of even parity. In its Weyl representation, it can be interpreted as the relativistic gravitational field of a finite, thin Newtonian rod with an arbitrary linear mass density. The ZV geometry thus provides a suitable seed for the application of the extended Fonarev scheme (see Appendix~\ref{appB} for a concise summary of the ZV geometry).

The asymptotically flat axisymmetric ZV spacetime \cite{Zipoy:1966btu,Voorhees:1970ywo}, also known as the $\gamma$-metric in the literature, in spherical coordinates is
\begin{equation}
\label{seed}
d s_{\mathrm{ZV}}^2=-f^\delta d t^2+ f^{-\delta}\left[\left(\frac{f}{g}\right)^{\delta^2} g\left(\frac{d r^2}{f}+r^2 d \theta^2\right)+f r^2 \sin ^2 \theta d \varphi^2\right],
\end{equation}
where the metric functions $(f,g)$ are given by 
\begin{align}
f=\left(1-\frac{2 M}{r}\right), \quad g=\left(1-\frac{2 M}{r}+\frac{M^2 \sin ^2 \theta}{r^2}\right). \label{ZVfunctions}
\end{align}
This metric is a vacuum solution of GR that deviates from the vacuum Schwarzschild black hole through a quadrupole deformation encoded in the deformation parameter $\delta$, such that for $\delta=1$, the geometry reduces to the Schwarzschild spacetime. Indeed, using Geroch’s definition of the multipole moments of an asymptotically flat gravitational field introduced in \cite{Geroch:1970cd}, it was shown in \cite{Quevedo:2012ttw, Abishev:2015pma} that the ZV solution is the simplest static, axially symmetric vacuum solution that possesses a quadrupole moment in addition to the mass parameter.
The two functions $f$ and $g$ vanish, respectively, at
\be
r_f = 2M, \qquad r^{\pm}_g = M (1 \pm \cos{\theta} ),
\ee
where one has $r^{+}_g \leqslant r_f$. Notice that at the equator, i.e. at $\theta=\pi/2$, the geometry reproduces the equatorial section of the spherically symmetric FJNW solution. In order to have a positive asymptotic ADM mass, one must impose $\delta >0$. 

The presence of the quadrupole modifies the point-like singularity structure of the Schwarzschild black hole. As discussed in detail in \cite{Kodama:2003ch}, the structure of the singularity depends on the range of the parameter $\delta$. For $ 0 <\delta <1 $, one has a string-like singularity, while for $\delta >1$, one has a ring-like singularity. In all cases, the singularity is located at $r=2M$ (with $\theta=\pi/2$ for the ring case), such that the solution is defined only for $2M < r < +\infty$.
Notice that, consequently, the zeroes of the function $g$ never manifest in this solution since $r^{+}_g  \leqslant 2M$.
Finally, this geometry can contain degenerate and non-rotating horizons for some specific values of $\delta$. Being degenerate and non-rotating, they do not describe a black hole horizon; thus, their existence does not contradict the no-hair theorem, which states the uniqueness of the Kerr solution for asymptotically flat and axis-symmetric black holes in four-dimensional GR. Moreover, when a horizon forms, it coincides with the ring singularity at the equator and is therefore itself singular at those points.

We can now examine the new time-dependent solution constructed from this static seed solution using the extended Fonarev method.

\subsection{The dynamical axisymmetric solution}
 
Applying the extended Fonarev theorem (\ref{extendedfonarev}) on this seed solution, one obtains an exact time-dependent extension of the Zipoy-Voorhees-FJNW solution given by 
\begin{align}
\label{METSOL}
d \tilde{s}^2 & =a^{2}(t)\left\{-f^{\delta \beta} d t^2+f^{-\delta \beta} \left[\left(\frac{f}{g}\right)^{\delta^2} g\left(\frac{d r^2}{f}+r^2 d \theta^2\right)+f r^2 \sin^2 \theta d \varphi^2\right]\right\}, \\
\label{SCALSOL}
\tilde{\phi} & =\delta \xi_0 \ln \left(1-\frac{2 M}{r}\right)+\frac{\xi_1\xi_2}{\kappa}\ln(Ct) ,
\end{align}
where the conformal factor reads
\be
\label{SCSOL}
a(t) = (C t)^{\xi_2}.
\ee
The time dependence of the solution therefore requires $\xi_2 \neq 0$, which corresponds to $\beta \neq \pm 1$. Indeed, dynamical behavior is possible only in the presence of scalar hair, which vanishes for $\beta = \pm 1$.
The coefficients $(\xi_0, \xi_1, \xi_2)$ are related to the parameters $(M, \delta, \beta)$ through \eqref{coeff}, while $C$ is a free real parameter.
To study the allowed coordinate ranges, let us first analyse the curvature singularities. These can be identified by examining the Ricci scalar, as the solution involves a scalar field in vacuum (see Appendix~\ref{appB}). Additionally, the time dependence of the solution generates a new spacelike singularity at $t = 0$, absent in the static case. Since $C$ is free, the solution naturally splits into two sectors depending on the sign of $C$. For the metric to remain Lorentzian, the allowed ranges of the radial and time coordinates are given by
\begin{align}
2M < r < +\infty \qquad \text{and} \qquad 
\left\{
    \begin{array}{ll}
        0 < t < +\infty   & \mbox{for } \; C > 0, \\
         - \infty < t < 0   & \mbox{for} \; C < 0 .
    \end{array}
\right.
\end{align}
One can switch between the two sectors through a standard time reversal of the geometry. This corresponds to the usual transformation between asymptotically FLRW black hole and white hole solutions. 

\subsubsection{Asymptotic FLRW behavior}

Let us first examine the properties of the solution in the large-$r$ time-dependent region. At sufficiently large $r$, one has $f(r) \sim 1$ and $g(r,\theta) \sim 1$, so that the solution reduces to a simple FLRW metric of the form
\begin{align}
d \tilde{s}^2 & =a^2(t)\left\{- d t^2+ d r^2+r^2 d \theta^2+r^2 \sin^2 \theta d \varphi^2 \right\},
\end{align}
filled with a homogeneous, time-dependent scalar field with profile
\begin{align}
\label{scal}
\tilde{\phi} (t)& =\frac{\xi_1\xi_2}{\kappa}\ln(Ct).
\end{align}
The scale factor and the corresponding conformal Hubble function are given by
\be
\label{scalefactor}
a(t) = (Ct)^{\xi_2}, \qquad \cH = \frac{\dot{a}}{a} = \frac{\xi_2}{t}.
\ee
The expanding and contracting branches depend on the signs of the constants $C$ and $\xi_2$. For $C > 0$ and $0 < t < +\infty$, the expanding (resp. contracting) branch corresponds to $\xi_2 > 0$ (resp. $\xi_2 < 0$). For $C < 0$ and $-\infty < t < 0$, the expanding (resp. contracting) branch corresponds to $\xi_2 < 0$ (resp. $\xi_2 > 0$). To illustrate the utility of the MCV, let us compute the position of the apparent horizon in the asymptotic FLRW geometry.
The mean curvature vector and its dual are expressed as
\begin{align}
H^{\mu} \partial_{\mu} = - \frac{1}{a^2} \left( \frac{1}{r}\partial_r + \cH \partial_t \right), \qquad H_{\perp}^{\mu} \partial_{\mu} = -  \frac{1}{a^2} \left( \cH \partial_r +\frac{1}{r} \partial_t \right).
\end{align}
The norm of the former reads
\be
H_{\mu} H^{\mu} = - H^{\perp}_{\mu} H_{\perp}^{\mu}= \frac{1}{a^2} \left( \frac{1}{r^2} - \cH^2\right)
\ee
By construction, this vector is null on the trapped surfaces of the geometry. Focusing on the solution with positive $r$, one recovers the standard position of the cosmological time-dependent horizon in a flat FLRW geometry, given by
\be
\label{cosmoapp}
r_{h}(t) = \cH^{-1}(t), 
\ee
or in comoving coordinates, i.e., $\tilde{r} = a(t) r$, where one has $\tilde{r} = \cH / a$. This is the unique trapped surface in this asymptotic region, as expected. We can further analyze the causal nature of this horizon. This can be determined by computing the ratio $\alpha = t'_{\text{CH}} / t'_{\text{LR}}$ between the velocity of the horizon trajectory $t'_{\text{CH}}$ and that of the radial light rays $t'_{\text{LR}}$. Imposing $\rd \theta = \rd \phi = 0$, the radial light rays satisfy $\rd s^2 = 0$, giving $t'_{\text{LR}} = \pm 1$, where the sign $\pm$ refers to outgoing or ingoing radial photons.
Focusing on the outgoing photons, the ratio is given by
\begin{align}
\alpha =\frac{t'_{\text{CH}}}{t'_{\text{LR}}} = -  \frac{ \xi_2}{r^2} .
\end{align}
Therefore, the horizon is never null. It is spacelike when $\xi_2 < 0$, which corresponds to $-1/\sqrt{2} < \beta < 1/ \sqrt{2}$, while it is timelike when $\xi_2 > 0$, i.e. for $ - 1 < \beta <  -1/\sqrt{2}$ or $ 1/\sqrt{2} < \beta < 1$.

\subsubsection{Identifying the time-dependent trapping horizons}

It is now natural to investigate the horizon structure of our dynamical ZV-FJNW solution. A natural starting point for such an analysis is, first, to construct the MCV and compute its norm, whose vanishing defines the boundary between (anti-)trapped and untrapped regions. Second, one should compute the corresponding expansions associated with the outgoing and ingoing null normals to the candidate horizon surface $\mathcal{S}$, and analyse their kinematical properties. 

In the case of spherical symmetry, this procedure is straightforward, as there is a natural choice for the embedding of $\mathcal{S}$ into the four-dimensional spacetime $\mathcal{M}$, namely the surface defined by constant $t$ and $r$, which is especially convenient in symmetric settings because it simplifies the computation and mimics spherical geometry.
In a generic dynamical axisymmetric spacetime, however, this choice is too restrictive: the geometry varies with the polar angle $\theta$, so the norm of the MCV may become a function $H^2(t,r,\theta)$, and then the condition $H^2(t,r,\theta)=0$, which determines the marginally trapped boundary, will typically be satisfied only for specific values of $\theta$, meaning that the surface is marginal only along certain angular loci rather than everywhere. 
This reveals a limitation for the canonical and simplest embedding: it can still detect the presence of marginally trapped behaviour locally in $\theta$, serving as a diagnostic that a true marginally trapped surface exists nearby, but it cannot represent it globally.

\vspace{5cm}
The appropriate generalisation is therefore to consider axisymmetric but deformed embeddings of the form $t=\mathrm{const}$ and $r=r(\theta)$, which allow the surface to adapt to the angular dependence of the geometry and yield a proper condition for a fully marginal (or trapped) surface through an equation for $r(\theta)$.

This is precisely what occurs with our dynamical ZV-FJNW spacetime. In fact, following the naive embedding, in our case, the metric on $\mathcal{S}$ takes the form
\be
\rd s^2_{\cS^2} = a^2r^2  f^{1-\delta \beta}\left[\left(\frac{g}{f}\right)^{1-\delta^2}  d \theta^2 + \sin ^2 \theta d \varphi^2\right],
\ee
where $a=a(t)$ is defined in (\ref{scalefactor}) and hence has the normal vectors 
\begin{align}
n_{\mu} \rd x^{\mu} = - a f^{\frac{\delta \beta}{2}} \rd t, \qquad s_{\mu} \rd x^{\mu} = a f^{\frac{\delta^2 - 1 - \delta\beta}{2}} g^{\frac{1-\delta^2}{2}} \rd r, \\
n^{\mu} \partial_{\mu} = a^{-1} f^{-\frac{\delta \beta}{2}} \partial_t,  \qquad s^{\mu} \partial_{\mu} = a^{-1} f^{\frac{1 + \delta\beta - \delta^2 }{2}} g^{\frac{\delta^2-1}{2}} \partial_r ,
\end{align}
with associated extrinsic curvatures given by 
\begin{align}
  K(n) & = \frac{2}{a}\mathcal{H}f^{-\frac{\delta\beta}{2}},\\
  K(s) & = -\frac{f^{\frac{1+\delta\beta-\delta^2}{2}}g^{\frac{1-\delta^2}{2}}}{a} \left[ \frac{2}{r} + \frac{\delta^2-2\delta\beta +1}{2}\partial_r \ln{f} + \frac{1-\delta^2}{2}\partial_r\ln{g} \right].
\end{align}
The norm of the MCV is explicitly found to be
\begin{equation}
  H^\mu H_\mu  = \frac{1}{4a^2f^{\delta\beta}} \left[ \mathcal{R}^2(r,\theta) - 4\mathcal{H}^2\right],
 \end{equation}
 where we have defined the function $R(r,\theta)$ as
 \begin{equation}
   \mathcal{R}(r,\theta) = \frac{f^{\frac{2\delta\beta - \delta^2 - 1}{2}} g^{\frac{\delta^2 - 3}{2}}}{r^3} \left[ 2 \left( r-2M \right)rg + M \left( 1-2\delta\beta + \delta^2 \right)rg + M \left( 1-\delta^2 \right) \left( r-M\sin{\theta}^2 \right)f \right].\label{R}
  \end{equation}
The locus at which the norm of the MCV vanishes depends non-trivially on the polar coordinate, rendering our initial embedding limited. 
A full characterisation of the causal structure of our time-dependent ZV-FJNW spacetime, therefore, requires a numerical treatment based on a more general embedding. Such an analysis is beyond the scope of the present work and is left for future investigation; however, we discuss its possible outcome in our conclusions. 

Nonetheless, before concluding this discussion, it is worth emphasising that there exist special cases in which the trivial embedding still provides a correct characterisation of the causal structure of a dynamical, axisymmetric spacetime. In such situations, the usefulness of the MCV is not diminished. This occurs for geometries in which the axisymmetric dependence of the norm of the MCV (and of the associated expansions) factorises, taking a generic form
\begin{equation}
H^2=\Theta(\theta)\bar{H}^2(t,r),
\end{equation}
with $\Theta(\theta)\neq0$ so that a curve $t=t(r)$, obtained from the initial embedding with $r=cte$, remains globally valid.
To date, the only known example of such a geometry has only recently been constructed in \cite{BenAchour:2026zui}, where it corresponds to a dynamically magnetised generalisation of the HMN solution. In that case, although the dynamical external magnetic field introduces a nontrivial axisymmetry into the otherwise spherically symmetric HMN spacetime, the horizon embedding $r=\mathrm{cte}$ remains consistent. As a result, the structure of the dynamical horizons can still be analyzed analytically by identifying the curve $t=t(r)$ along which the norm of the MCV vanishes.

\section{Discussion}

In this work, we have provided two new results to advance the description of axisymmetric compact objects embedded in cosmology. First, we have introduced a new solution-generating technique that allows us to derive the first asymptotically FLRW and axisymmetric inhomogeneous family of solutions in the self-interacting Einstein-scalar system. Using this method, we have constructed a first analytic solution of this kind and sketched some of its features.
In parallel, we reviewed and showed explicitly on a concrete example that the mean curvature vector introduced by Anco in \cite{Anco:2004bb} constitutes the natural generalisation of the Kodama vector beyond spherical symmetry. This generalisation enables the invariant identification—via the norm of the vector—of (anti)-trapped regions in any geometry, without requiring specific symmetry, although some embedding dependence still prevails, providing limitations to be circumvented numerically for generic dynamical and axisymmetric spacetimes.  

Let us now discuss these results in more detail.

\begin{itemize}
\item \textit{Extension of the Fonarev method for axisymmetric dynamical solutions:} The new solution-generating technique summarized in the theorem (\ref{extendedfonarev}) represents a direct extension of the method developed by Fonarev in \cite{Fonarev:1994xq} in the spherically symmetric context. We emphasize that this method allows the construction of an entire new family of asymptotically FLRW and axisymmetric compact objects. The new time-dependent Zipoy-Voorhees solution (\ref{METSOL} - \ref{SCSOL}) is only one example among many.
The core idea of the method is to combine: i) the original Fonarev approach \cite{Fonarev:1994xq}, which constructs non-stationary spherically symmetric scalar-vacuum solutions, and ii) the extended Buchdahl method studied in \cite{Barrientos:2024uuq}, which produces stationary axisymmetric scalar-vacuum solutions. 
\item \textit{Difference with the conformally transformed solutions and the role of the self-interaction:} It is important to emphasize that the extended Fonarev method differs fundamentally from the standard techniques based on conformal transformations of vacuum black hole solutions, such as those used to derive the Thakurta metric. Indeed, the latter approach consists of conformally transforming a vacuum black hole solution (e.g., Kerr) with a time-dependent scale factor and then using the Einstein equations to determine the associated energy-momentum tensor $T_{\mu\nu}$. While this procedure is rather economical, it does not guarantee a matter source free of pathologies, as exemplified by the Thakurta solution.
In contrast, the Fonarev method introduces both a conformal rescaling of the seed metric and a time-dependent shift of the stationary scalar profile. These two functions are then solved exactly via the (reduced) Einstein field equations. In this construction, the form of the Liouville potential in the target theory is crucial, as it allows the time-dependent scale factor to be balanced when solving the equations. Except for the special case of the HMN solution and its new axisymmetric extension, where the potential vanishes, the non-stationarity of the new solutions is intimately tied to the presence of the self-interacting potential.
\item \textit{Properties of the new time-dependent Zipoy-Voorhees solution:} The new solution derived from our solution-generating technique describes a non-vacuum asymptotically FLRW compact object. It is characterized by four parameters: the mass $M$, the scalar charge $\beta$, and the parameters $(\delta, C)$.
Asymptotically, the geometry corresponds to a flat FLRW cosmology filled with a time-dependent scalar field. Thanks to the self-interacting potential, the equation of state parameter $\omega$ can be adjusted to reproduce all possible dynamics. It is important to emphasize that, in this case, the full machinery of the MCV is not sufficient to completely characterize (without numerical tools) the horizon structure of such dynamical, axisymmetric spacetimes, and that numerical methods must be implemented. Nevertheless, this geometry provides valuable insight into the extent to which the MCV can yield exact results, and how it may still be effectively applied to other dynamical, axisymmetric solutions, such as those recently reported in \cite{BenAchour:2026zui}.
\item\textit{Formation of the dynamical horizons:} 
The formation of dynamical horizons in the present dynamical ZV--FJNW solution remains to be clarified, not due to a conceptual difficulty but because, in this work, we have not carried out the numerical construction of a suitable embedding that would determine the horizon curve in full generality. Nevertheless, the possible outcomes are relatively clear. Either the horizon structure is genuinely different from that of the HMN or Fonarev spherically symmetric solutions, in which case little can be concluded without a dedicated analysis, or the horizon curve follows the so-called S-curve behavior typically found in dynamical Einstein-scalar systems \cite{Faraoni:2015ula}.

In the latter case, one expects the equations governing the horizon trajectories to show that cosmological horizons and black (or white) hole horizons appear and annihilate in pairs. More specifically, starting from the initial big-bang singularity at time $t_0 = 0$ up to a first critical time $t_1$, the spacetime would contain a single, expanding cosmological horizon. Shortly after $t_1$, two additional horizons would form: one corresponding to another cosmological horizon that continues to expand, and the other to a contracting horizon associated with the compact object. This latter horizon would shrink over time until it merges with the original cosmological horizon at a second critical time $t_2$. For $t > t_2$, only the later-formed cosmological horizon would remain, leaving a naked singularity embedded in an FLRW background.

Within this scenario, the horizon associated with the compact object would form at the same location as the cosmological horizon. As a result, the solution would acquire physical relevance only after the critical time of horizon formation, when it can describe a compact object, potentially a black hole, embedded in a cosmological spacetime.

This interpretation, however, should be treated with caution, as it is not unambiguous. Due to the genuinely dynamical nature of the solution, no eternal event horizon forms, and the spacetime does not contain a black hole in the global sense. At most, a transient dynamical horizon develops, covering the singularity only for a finite interval and corresponding to a temporary trapped region. From this perspective, the solution may be better viewed as a cosmological spacetime with a central singularity whose ability to trap light is not permanent.

Although this type of behavior has been observed in exact dynamical solutions, its underlying mechanism is still not fully understood and deserves further study. In particular, it would be valuable to compare these features with results from numerical simulations of gravitational collapse.

\end{itemize}
Our study opens several directions for further investigation. A first natural question is whether the solution-generating method can be generalised to other classes of scalar field potentials, or extended to include rotation\footnote{Rotation is particularly challenging, as we have shown here. Nevertheless, a magnetised version of this class of spacetimes has been constructed recently, yielding solutions that are naturally axisymmetric \cite{BenAchour:2026zui}.}. Established techniques based on higher-dimensional embeddings suggest that such rotating dynamical solutions may indeed be achievable. More broadly, the proposed method provides a starting point for exploring new regions of the solution space of the Einstein–Scalar field system describing non-spherical compact objects embedded in cosmology. This may be relevant for (i) extending and testing geometrical tools developed in spherical symmetry, (ii) studying the phenomenology of PBHs, and (iii) providing a model of scalar accretion onto a black hole. Indeed, a cosmological contracting branch for the type of solutions presented here can also be used to model a given sub-region of spacetime where a compact object is accreting a scalar field, the contraction being triggered by this accretion process. Moreover, it would be of interest to employ one of these new background solutions to investigate semi-classical Hawking evaporation within a fully dynamical geometry, using, for instance, the dynamical notion of temperature for general dynamical horizons introduced in \cite{Senovilla:2014ika}. This could allow us to confront the current constraints on the PBH mass range based on the semi-classical treatment of the stationary asymptotically flat black holes.

\begin{appendices}

\section{Appendix A}
\label{appA}

As stated above, the Buchdahl and Fonarev theorems can be effectively combined to construct dynamical axisymmetric exact solutions starting from a static axisymmetric seed. A natural question that arises is whether it is possible to extend this construction to dynamical rotating solutions. A first step in this direction would be to devise a mechanism for generating stationary rotating solutions with scalar hair. This has been achieved thanks to the Eriş–Gürses theorem \cite{Eris:1976xj}, and has been extensively discussed and explored in \cite{Barrientos:2025abs}. However, it appears that the Eriş–Gürses theorem does not admit a straightforward unification with the Fonarev technique. One way to see an immediate obstruction is to directly consider the following rotating configuration with scalar hair
\begin{align}
g_{\mu \nu} & =e^{2 \mu(t)} {g}_{\mu \nu}(r, \theta), \\
\phi & ={\phi}(r, \theta)+\xi_1\Psi(t),
\end{align}
and to compute the field equations. Here, ${g}_{\mu\nu}$ is a rotating metric with a scalar hair profile sourced by ${\phi}(r,\theta)$.
After some algebra, the system reduces to 
\begin{subequations}
\begin{align}
-3 \ddot{\mu} & =\kappa\left(\xi_1^2 \dot{\Psi}^2+e^{2 \mu} {g}_{t t} V\right), \\
\dot{\mu}\left[{g}^{t t} \partial_r {g}_{t t}+{g}^{t \varphi} \partial_r {g}_{t \varphi}\right] & =\kappa  \xi_1 \dot{\Psi} \partial_r {\phi}, \\
\dot{\mu}\left[{g}^{t t} \partial_\theta {g}_{t t}+{g}^{t \varphi} \partial_\theta {g}_{t \varphi}\right] & =\kappa  \xi_1 \dot{\Psi} \partial_\theta {\phi}, \\
-{g}^{t t}\left(\ddot{\mu}+2 \dot{\mu}^2\right) & =\kappa e^{2 \mu} V, \\
\dot{\mu}\left[{g}^{t t} \partial_r {g}_{t \varphi}+{g}^{t \varphi} \partial_r {g}_{\varphi \varphi}\right] & =0, \\
\dot{\mu}\left[{g}^{t t} \partial_\theta {g}_{t \varphi}+{g}^{t \varphi} \partial_\theta {g}_{\varphi \varphi}\right] & =0 .
\end{align}
\end{subequations}
In order to obtain a nontrivial conformal factor $\mu$, then an initial constraint is required
\begin{subequations}
\label{nogoeq}
\begin{align}
{g}^{t t} \partial_r {g}_{t \varphi}+{g}^{t \varphi} \partial_r {g}_{\varphi \varphi}&=0,\\
{g}^{t t} \partial_\theta {g}_{t \varphi}+{g}^{t \varphi} \partial_\theta {g}_{\varphi \varphi}&=0. 
\end{align}
\end{subequations}
It can be straightforwardly shown that these conditions cannot be satisfied. The reasoning is as follows. The Eriş–Gürses theorem operates in such a way that the entire effect of the scalar field profile is absorbed into a modification of the non-Killing sector of the metric. Consequently, a Kerr spacetime with scalar hair—such as the one presented in \cite{Barrientos:2025abs} or any other solution obtained through the Eriş–Gürses method—necessarily retains the same Killing components as the standard Kerr geometry. Only the $rr$ and $\theta\theta$ components experience the backreaction from the scalar field. As a result, the condition in question reduces to a requirement on the Kerr metric components themselves. It is immediate to verify that this requirement is not met by the Kerr line element.

An alternative approach would be to extend Buchdahl’s theorem to include rotation, at least within a perturbative, slowly rotating framework. While this is certainly feasible, the metric components obtained at first order in the rotational parameter do not satisfy \eqref{nogoeq}. In fact, the constraint \eqref{nogoeq} enforces a specific relation between the metric components $g_{t\varphi}$ and $g_{\varphi\varphi}$,
\begin{equation}
g_{t\varphi}=\Omega_0 g_{\varphi\varphi},
\end{equation}
with $\Omega_0$ a constant. This condition implies that the only form of rotation permitted by the system is that obtained through a transformation of the form $\varphi\rightarrow\varphi-\Omega_0 t$, which is merely a change of reference frame rather than a physically distinct rotating solution.

\section{Appendix B}
\label{appB}

The ZV spacetime \cite{Zipoy:1966btu,Voorhees:1970ywo} is a static and axisymmetric, asymptotically flat solution of the vacuum Einstein equations. It represents an axisymmetric extension of the Schwarzschild geometry that incorporates a quadrupole moment in addition to the mass parameter. 

General static and axisymmetric spacetimes are described by the Weyl metric, which in canonical coordinates $-\infty<t<\infty$, $0\leq\rho<\infty$, $-\infty<z<\infty$ and $0\leq\varphi<2\pi$ reads
\begin{equation}
d s^2=-e^{2\Phi} d t^2+e^{-2\Phi}\left[e^{2 \gamma}\left(d \rho^2+d z^2\right)+\rho^2 d \phi^2\right],
\end{equation}
where the Newtonian potential $\Phi$ and the metric component in front of the non-Killing sector of the metric, $\gamma$, are functions of $\rho$ and $z$ only.
It is well known that the entire spectrum of static and axisymmetric vacuum solutions is mathematically determined, since the potential $\Phi$ satisfies a Laplace equation in three-dimensional Euclidean space with cylindrical coordinates,
\begin{equation}
\frac{\partial^2\Phi}{\partial\rho^2}+\frac{1}{\rho}\frac{\partial\Phi}{\partial\rho}+ \frac{\partial^2\Phi}{\partial z^2}=0,
\end{equation}
which is the reason why $\Phi$ is treated as a Newtonian potential. The remaining function $\gamma$ can then be integrated through the quadratures
\begin{equation}
\frac{\partial\gamma}{\partial\rho}=\rho\left[ \left(\frac{\partial\Phi}{\partial\rho}\right)^2+\left(\frac{\partial\Phi}{\partial z}\right)^2\right], \qquad \frac{\partial\gamma}{\partial z}=2\rho \left(\frac{\partial\Phi}{\partial\rho}\right)\left(\frac{\partial\Phi}{\partial z}\right).
\end{equation}
The Laplace equation admits a full solution in terms of a multipolar expansion, with terms that decay far from the source (ensuring asymptotic flatness) and terms that decay when approaching the source, terms that diverge asymptotically.
Consequently, $\gamma$ can always be determined. This establishes a nontrivial connection between Newtonian sources and relativistic line elements, as $\Phi$ effectively represents the Newtonian gravitational field.

The ZV metric can be interpreted as the relativistic analogue of the Newtonian gravitational potential produced by a thin rod of mass $m$ and length $2l$. The metric functions are given by
\begin{equation}
\mathrm{e}^{2 \Phi}=\left(\frac{R_{+}+R_{-}-2 l}{R_{+}+R_{-}+2 l}\right)^\delta, \qquad \mathrm{e}^{2 \gamma}=\left[\frac{\left(R_{+}+R_{-}\right)^2-4 l^2}{4 R_{+} R_{-}}\right]^{\delta^2},
\end{equation}
where $\delta=m/l$ is a constant parameter and $R_{ \pm}:=\sqrt{\rho^2+(z \pm l)^2}$. The solution is obtained from the asymptotically flat sector of the multipolar expansion of $\Phi$, in which only the parity-even terms contribute, with multipolar coefficients fixed as $a_n=ml^n/(n+1)$ \cite{Katsumata:2025jrf}.
The Minkowski background is recovered for $\delta=0$, while the Schwarzschild black hole is obtained when $\delta=1$. The parameter $\delta$ is known as the deformation parameter: for $0<\delta<1$ the spacetime is oblate, while for $\delta>1$ it is prolate.

To understand the singularity structure of the ZV geometry, it is convenient to use the so-called prolate spheroidal coordinates $(x,y)$ that connect with $(\rho,z)$ via 
\begin{equation}
\rho=l \sqrt{\left(x^2-1\right)\left(1-y^2\right)}, \quad z=l x y,
\end{equation}
and for which the ZV line element takes the form 
\begin{equation}
d s^2=-e^{2\Phi} d t^2+\Sigma^2\left(\frac{d x^2}{x^2-1}+\frac{d y^2}{1-y^2}\right)+R^2 d \phi^2, 
\end{equation}
where 
\begin{subequations}
\begin{align}
    e^{2\Phi}&=\left(\frac{x-1}{x+1}\right)^\delta,\\
    \Sigma^2&=l^2 \frac{(x+1)^{\delta(1+\delta)}}{(x-1)^{\delta(1-\delta)}}\left(x^2-y^2\right)^{1-\delta^2},\\
    R^2&=l^2\left(\frac{x+1}{x-1}\right)^{\delta-1}(x+1)^2\left(1-y^2\right).
\end{align}
\end{subequations}
A curvature singularity arises at a specific value of $x$. Accordingly (see below), the asymptotically flat sector of the spacetime is naturally restricted to $x>1$ with $-1 \leq y \leq 1$.
The asymptotic behavior of the Newtonian potential shows (when expressed in spherical-like coordinates) that the mass is given by $m=l\delta$. Therefore, it is natural to focus on the case where $\delta>0$. 
These coordinates are particularly convenient for analyzing the singularity structure of the spacetime. In fact, the Kretschmann invariant simplifies to
\begin{equation}
R_{\mu\nu\lambda\rho}R^{\mu\nu\lambda\rho}=\frac{4 (x^2 - y^2)^{-3 + 2\delta^2} (x - 1)^{-2 - 2\delta^2 + 2\delta} (x + 1)^{-2 - 2\delta^2 - 2\delta}}{m^4}F(x,y,\delta),
\end{equation}
with $F(x,y,\delta)$ being a lengthy polynomial expression free of poles.
Following the analysis of \cite{Kodama:2003ch}, one concludes that in the asymptotically flat region $x>1$ with $-1 \leq y \leq 1$, the spacetime exhibits a singularity along the open segment $\rho=0$, $-l \leq z \leq l$, which is equivalently described by $(x=1, -1 \leq y \leq 1)$, provided $\delta \neq 0,1$, values for which the Minkowski and Schwarzschild geometries are retrieved. The nature of the singularity has been studied in detail: it is point-like for $\delta < 0$, string-like for $0 < \delta < 1$, and ring-like for $\delta > 1$. Again, $\delta$ is restricted to be positive, ensuring a positive mass.

The extension of the ZV spacetime with a minimally coupled scalar field, as given by Buchdahl's theorem, is defined by the metric functions 
\begin{subequations}
\begin{align}
    e^{2\Phi}&=\left(\frac{x-1}{x+1}\right)^{\delta\beta},\\
    \Sigma^2&=l^2 \frac{(x+1)^{\delta^2+\beta\delta}}{(x-1)^{-\delta^2+\beta\delta}}\left(x^2-y^2\right)^{1-\delta^2},\\
    R^2&=l^2\left(\frac{x+1}{x-1}\right)^{\beta\delta-1}(x+1)^2\left(1-y^2\right).
\end{align}
\end{subequations}
and the scalar field profile 
\begin{equation}
\phi=\frac{\delta}{2}\sqrt{1-\beta^2}\ln\left(\frac{x-1}{x+1}\right). 
\end{equation}
In this case, the Kretschmann invariant takes the form
\begin{equation}
R_{\mu\nu\lambda\rho}R^{\mu\nu\lambda\rho}=\frac{4 (x^2 - y^2)^{-3 + 2\delta^2} (x - 1)^{-2 - 2\delta^2 + 2\beta\delta} (x + 1)^{-2 - 2\delta^2 - 2\beta\delta}}{m^4}\tilde{F}(x,y,\delta),
\end{equation}
where $\tilde{F}(x,y,\delta)$ denotes another polynomial expression free of poles. The qualitative structure of the singularity remains unchanged with respect to the vacuum case. 

Finally, using the Erez-Rosen coordinates $(r,\theta)$ 
\begin{equation}
    x=\frac{r}{m}-1, \qquad y=\cos\theta,
\end{equation}
the ZV-FJNW solution takes the form 
\begin{align}
d s_{\mathrm{ZV-FJNW}}^2 & =-f^{\delta \beta} d t^2+\frac{\left[\left(\frac{f}{g}\right)^{\delta^2} g\left(\frac{d r^2}{f}+r^2 d \theta^2\right)+f r^2 \sin ^2 \theta d \varphi^2\right]}{f^{\delta \beta}}, \\
\Phi & =\frac{\delta}{2} \sqrt{1-\beta^2} \ln \left(1-\frac{2 M}{r}\right),
\end{align}
with metric functions $f$ and $g$ given in \eqref{ZVfunctions}, from which the vacuum ZV case in these coordinates is recovered in the limit $\beta \rightarrow 1$. Notice that this non-dynamical spacetime is the one recovered if suppressing the time dependence of our novel configuration given in \eqref{METSOL} and \eqref{SCALSOL}.

\end{appendices}

\section*{Acknowledgments}
The authors gratefully acknowledge insightful discussions with Amaro D\'iaz, Jos\'e Barrientos and Keanu Muller. J. BA thanks Vincent Vennin for several discussions on primordial black holes during the first steps of this project and acknowledges the hospitality of the Yukawa Institute for Theoretical Physics during the workshop "Challenge in Gravity and Cosmology - 2023" where this project was initiated. The authors would also like to express their gratitude to the anonymous referee for her/his very insightful comments. 
A.C. is partially supported by FONDECYT grant 1250318. M.H gratefully acknowledges the University of Paris-Saclay for its warm hospitality during the development of this project.

\bibliography{apssamp.bib}

\bibliographystyle{unsrt}

\end{document}